\documentclass{aa}  

\usepackage{array}
\usepackage{graphicx}
\usepackage{txfonts}
\usepackage{multirow}
\usepackage{amsmath}
\usepackage{txfonts}
\usepackage{xcolor}
\usepackage{graphicx}
\usepackage{bm}
\usepackage{float}

\usepackage{natbib}
\usepackage{booktabs}
\usepackage{fixltx2e}

\usepackage{url}
\usepackage{hyperref}
\hypersetup{
   colorlinks=true,       
   linkcolor=black,       
   citecolor=blue,        
   urlcolor=black         
}

\newcommand{\HI}{\ifmmode \mathrm{\ion{H}{I}} \else \ion{H}{I} \fi}
\newcommand{\degree}[0]{$^{\circ}$}
\newcommand\ab{\bm{a}}
\newcommand\mub{\bm{\mu}}
\newcommand\sigmab{\bm{\sigma}}
\newcommand{\IHI}{\ifmmode I_{{\mathrm{H}} \, \mathrm{I}} \else $I_{{\mathrm{H}} \, \mathrm{I}}$\fi} 
\def\GHz{\ifmmode $\,GHz$\else \,GHz\fi}
\def\MJysr{\ifmmode \,$MJy\,sr\mo$\else \,MJy\,sr\mo\fi}
\def\microns{\ifmmode \,\mu$m$\else \,$\mu$m\fi}
\def\micron{\microns}
\def\kms{\ifmmode $\,km\,s$^{-1}\else \,km\,s$^{-1}$\fi}

\begin{document} 

   \title{Separation of dust emission from the cosmic infrared background in \textit{Herschel} observations with wavelet phase harmonics}
   
   \author{Constant Auclair \inst{1}, 
   Erwan Allys \inst{1}, 
   François Boulanger \inst{1}, \\
   Matthieu Béthermin \inst{2}, 
   Athanasia Gkogkou \inst{2},
   Guilaine Lagache \inst{2},
   Antoine Marchal \inst{3,4}, \\
   Marc-Antoine Miville-Deschênes \inst{5},
   Bruno Régaldo-Saint Blancard \inst{6},
   Pablo Richard \inst{1}}
    
   \institute{
   \inst{1} Laboratoire de Physique de l’École normale supérieure, ENS, Université PSL, CNRS, Sorbonne Université, Université Paris Cité, F-75005 Paris, France \\
    \inst{2} Aix Marseille Univ, CNRS, LAM, Laboratoire d’Astrophysique de Marseille, Marseille, France \\
    \inst{3} Canadian Institute for Theoretical Astrophysics, University of Toronto, 60 St. George Street, Toronto, ON M5S 3H8, Canada \\
    \inst{4} Research School of Astronomy \& Astrophysics, Autralian National University, Cambera, ACT, 2610 Australia \\
   \inst{5} Institut d'Astrophysique Spatiale, Université Paris-Sud, Bât. 121, 91405 Orsay, France \\
   \inst{6} Center for Computational Mathematics, Flatiron Institute, 162 5th Avenue, New York, NY 10010, USA
   }

   \date{Submitted to \aap\ on May 4, 2023 \slash \ Accepted October 23, 2023}
 
  \abstract{
   {The low-brightness dust emission at high Galactic latitudes is of interest with respect to studying the interplay among the physical processes involved in shaping the structure of the interstellar medium (ISM), as well as in statistical characterizations of the dust emission as a foreground to the cosmic microwave background (CMB). Progress in this avenue of research has been hampered by the difficulty related to separating the dust emission from the cosmic infrared background (CIB).}
   {We demonstrate that the dust and CIB may be effectively separated based on their different structure on the sky and we use the separation to characterize the structure of diffuse dust emission on angular scales, where the CIB is a significant component in terms of power.}
   {We used scattering transform statistics, wavelet phase harmonics (WPH) to perform a statistical component separation using \textit{Herschel} SPIRE observations. This component separation is done only from observational data using non-Gaussian properties as a lever arm and is done at a single $250\, \mu$m frequency. This method, which we validated on mock data, gives us access to non-Gaussian statistics of the interstellar dust and an output dust map that is essentially free from CIB contamination.}
   {Our statistical modeling characterizes the non-Gaussian structure of the diffuse ISM down to the smallest scales observed by Herschel. We recovered the power law shape of the dust power spectrum up to $k = 2 \ \text{arcmin}^{-1}$, where the dust signal represents $2 \%$ of the total power. Going beyond the standard power spectra analysis, we show that the non-Gaussian properties of the dust emission are not scale-invariant. The output dust map reveals coherent structures at the smallest scales, which had been hidden by the CIB anisotropies. This aspect opens up new observational perspectives on the formation of structure in the diffuse ISM, which we discuss here in reference to a previous work.}
   {We have succeeded in performing a statistical separation from the observational data at a single frequency by using non-Gaussian statistics. }}


   \authorrunning{Auclair et al.}

   \titlerunning{Separation of dust emission from the CIB in \textit{Herschel} observations with WPH}

   \maketitle

%

\section{Introduction}

Thermal emission from interstellar dust and the cosmic infrared background (CIB) are the two main emission components of the sky at far-infrared and sub-millimeter wavelengths. 
The emission from dust is a tracer of diffuse interstellar matter imaged for the first time across the sky by the IRAS space mission \citep{Gautier92,Mamd07}. More recently, the \textit{Herschel} space mission has provided data at a higher angular resolution, which are useful in characterizing the filamentary structure of molecular clouds \citep{Andre10,Mamd10,Robitaille19,Yahia21}. The CIB is the diffuse background emission associated with the dust emission from galaxies integrated over their cosmic evolution \citep{Hauser01}. It provides us with information on galaxy evolution and the large-scale structure of the universe \citep{Knox01,Bethermin13,PlanckXXX_CIB,Maniyar18}. Power spectra are the simplest means of analysing far-infrared maps of the sky and statistically distinguishing dust and CIB anisotropies. At high Galactic latitudes, the two sources of emission are entwined and the CIB power spectrum dominates that of the emission from diffuse interstellar matter on angular scales smaller than about $1^\circ$ \citep{Mamd02,Lagache07,Viero13,PlanckXXX_CIB,Mak17}. 

\begin{figure*}
    \centering
    \resizebox{\hsize}{!}{\includegraphics{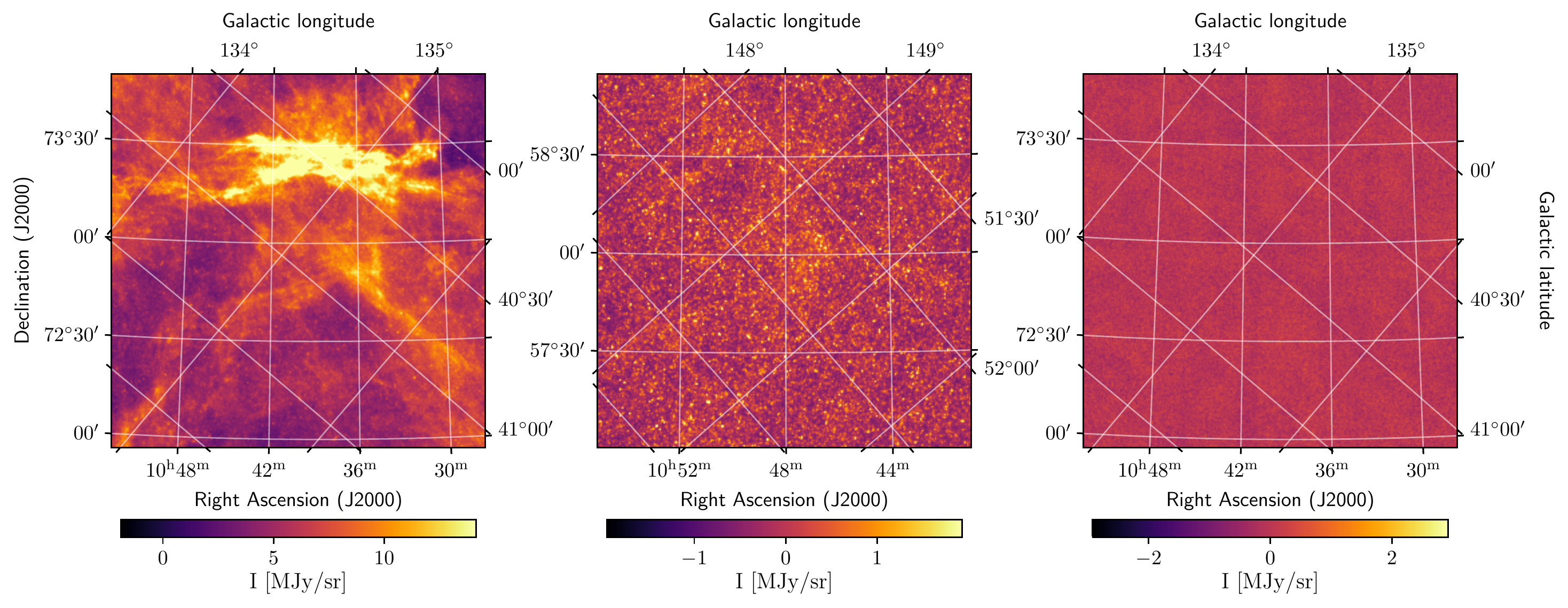}}
    \caption{\textit{Herschel} SPIRE maps at $250\, \mu$m. Left: Map of the Spider field, $d_S$. Center: Map of the LH field, $d_L$.  Right: Data noise for the Spider observations, $d_N$.}
    \label{fig:Herschel_observations}
\end{figure*} 

The separation of the dust emission from CIB anisotropies is a major difficulty, 
hampering the analysis of the low-brightness emission from the diffuse interstellar medium (ISM). Dust emission at high Galactic latitudes is of specific interest in studying the interplay between thermal instability, interstellar turbulence and magnetic fields in shaping the structure of the ISM \citep{Vazquez00,Kritsuk02,saury_structure_2014,Hennebelle19}, as well as to statistically characterize dust emission foreground to the cosmic microwave background (CMB) \citep{Jewell01,Mamd07}.  

Several component separation methods were employed to analyze the {\it Planck} sky maps. 
They were based on combinations of frequency maps or modeling of the spectral energy distribution (SED) \citep{Planck2018_IV}. To separate the dust emission from the CIB, we need to follow a different approach using the structure on the sky because the two components have very similar SEDs. The generalized needlet internal linear combination (GNILC) method has led the way in this regard. \citet{PIPXLVIII} produced dust emission maps that are corrected for a large part of the CIB anisotropies, but at the expense of losing the structure of the dust emission on small angular scales, where the CIB is dominant, as well as of keeping residual CIB emission on large angular scales where the dust is dominant~\citep{chiang_extragalactic_2019}. The correlation between dust and \HI\ emission has been extensively used to produce CIB maps but this method does not yield a dust map independent of \HI\ \citep{boulanger_dustgas_1996,PlanckXXX_CIB,lenz_large-scale_2019}. 

This paper circumvents these limitations by using scattering transforms \citep{mallat_group_2012,bruna2013invariant}.
These statistics are similar to convolution neural networks but are written in an explicit mathematical form. They combine convolutions of the input image with wavelets on several oriented scales, with a non-linear operator to capture interactions between scales as specific imprints of non-Gaussian textures \citep{Cheng21}. Based on predefined wavelet filters, these summary statistics can be used to characterize the non-Gaussian texture of images without any learning step. 
We make use of two variants of  scattering transforms: the wavelet scattering transform (WST) and the wavelet phase harmonics (WPH). The WST has been used to analyze synthetic maps of dust total intensity and polarization built from MHD simulations of interstellar clouds in \citet{allys_rwst_2019}, \citet{regaldo-saint_blancard_statistical_2020}, and \cite{Saydjari21}. \citet{allys_rwst_2019} also presented a first application to a \textit{Herschel} observation of dust emission. The WPH statistics were introduced in cosmology to analyze simulations of the large-scale structure of the Universe by \citet{allys_new_2020} who showed that they may be used to synthesize maps reproducing the non-Gaussian texture of the input image. \citet{regaldo-saint_blancard_new_2021} used this capability to statistically separate dust emission from data noise in \textit{Planck} polarization maps, using their different non-Gaussian properties. This paper extends this approach to the separation of dust and CIB emission using \textit{Herschel} SPIRE observations at a single $250\, \mu$m wavelength, performing a statistical component separation relying solely on observational data. Our scientific motivation is twofold. 
First, we want to demonstrate that dust and CIB may be effectively separated based on their different structure on the sky. Second, we want to use the separation to characterize the structure of  diffuse dust emission at high Galactic latitude on angular scales where the CIB is the dominant component in terms of power. 

The paper is organized as follows. 
We present the observations we use in Sect.~\ref{sec:obs}. Our component separation method is introduced in Sect.~\ref{sec:comp_sep} and validated on mock data in Sect.~\ref{sec:validation}. In Sect.~\ref{sec:application} we present the results of our dust and CIB component separation based on \textit{Herschel} SPIRE observations. The output dust map is used to characterize the non-Gaussian structure of diffuse dust emission at high Galactic latitude in Sect.~\ref{sec:non-Gaussian analysis}.
The paper results and perspectives are summarized in Sect.~\ref{sec:summary}. The paper has five appendices that present a summary of the mathematical notations used in this paper (Appendix~\ref{app:notations}), the specific set of WPH statistics used in the paper (Appendix~\ref{app:WPH_statistics}), the \HI data used to build the mock data (Appendix~\ref{app:HI_map}), the WPH statistics of dust emission for the mock and \textit{Herschel} data (Appendix~\ref{app:WPH_validation}), and a brief presentation of the reduced wavelet scattering transform (RWST, Appendix~\ref{app:RWST}).

\section{Observations}
\label{sec:obs}

This study makes use of observations obtained with the Spectral and Photometric Imaging REceiver (SPIRE) on the {\it Herschel} Space Observatory \citep{Griffin10}. The SPIRE photometer was used to image the sky emission in three broad spectral bands centred at 250, 350, and 500\micron. We only use the 250\micron\ (corresponding to a frequency of 1200 GHz) images, which provide an angular resolution of $18^{\prime\prime}$ (full width at half maximum, FWHM, of the beam). 

We aim to separate the dust and CIB emissions in an observation of diffuse interstellar matter in the so-called Spider region at a high Galactic latitude. To characterize the CIB statistics, we also made use of a SPIRE observation towards the Lockman Hole (LH), a sky area known to have a small amount of interstellar matter from \HI 21\,cm observations \citep{Lockman86}. The Spider observations are presented in Sect.~\ref{subsec:Spider} and the LH one is given in Sect.~\ref{subsec:LH}.

\subsection{Spider region}
\label{subsec:Spider}

The Spider field \citep[named for its prominent "legs" emanating from a central ``body,"][and references within]{Marchal23} targets diffuse interstellar matter at high Galactic latitudes
along a segment of the North Celestial Pole (NCP) Loop \citep{heiles_hi_1984,heiles_magnetic_1989,meyer_discovery_1991}, a conspicuous feature of \HI and dust emission extending over $\sim 30^\circ$ in the northern sky.
The SPIRE observations of this field were performed in the same way as described by \citet{Mamd10} for the Polaris field. The Spider field was observed twice in different directions. 

To have two independent maps (one for each scanning direction) with distinct noise realizations, we use maps from the Level~2 stage of the data processing. These maps are combined at the Level~3 stage of the data processing.  We checked that the Level~2 and 3 maps of the Spider region differ only by a signal compatible with instrumental noise.
The maps were downloaded from The \textit{Herschel} Archive\footnote{The file names are $\rm{hspirepsw1342231359\_20pxmp\_1462623735063}$ and $\rm{hspirepsw1342231360\_20pxmp\_1462624580145}$.}.  
We refer to the start guide to {\it Herschel}-SPIRE\footnote{\url{https://www.cosmos.esa.int/documents/12133/1035800/QUICK-START+GUIDE+TO+HERSCHEL-SPIRE}} for the data processing and map characteristics. 

Here, we refer to the two Spider maps as $d_{S1}$ and $d_{S2}$.
These maps contain very bright galaxies that have a strong influence on our statistical analysis. We removed the $\sim 70$ bright sources whose peak brightness is over 6 MJy/sr by replacing the corresponding pixels by a close similar area of the same map. We identified these sources directly from the maps, but we have verified that they are all extragalactic. We also checked that our results are not particularly sensitive to the way we chose to remove the sources. Then, we can define:
\begin{equation}
    \displaystyle d_S = \frac{d_{S1} + d_{S2}}{2}, \ \ d_N = \frac{d_{S1} - d_{S2}}{2}.
    \label{eq:d_S}
\end{equation}
The mean map, $d_S$, images the total (dust+CIB) noisy infrared emission. We take the half-difference map, $d_N$, as a statistical realization of the instrumental noise of $d_S$. We note that this map does not include instrument systematics that would have exactly the same imprint in both $d_{S1}$ and $d_{S2}$. The two images are displayed in Fig.~\ref{fig:Herschel_observations}. These are squares with sides of 1144 pixels corresponding to 1.91 degrees on the sky\footnote{To apply our algorithm to non-periodic maps requires to use additional pixels around their edges, that cannot be used for the subsequent analysis. In this paper, we only describe the central part of the map on which the analysis is performed. For instance, the component separation of the $d_S$ map involves a larger map of 1400x1400 pixels, but we describe here only the corresponding 1144x1144 central region.}. The pixel size of the images is $6^{\prime\prime}$.

\subsection{LH region}
\label{subsec:LH}

LH is one of the fields targeted by the \textit{Herschel} Multi-tiered Extragalactic Survey \citep[HERMES,][]{Oliver12}. \citet{Viero13} have used the data to characterize the structure of the CIB with the power spectra. Their analysis shows that on this field the contamination of the extended emission by Galactic dust is much smaller than in the Spider field. We have used the Level~3 image.

The LH map at 250\micron\ was downloaded from the \textit{Herschel} Database in Marseille \citep[HeDaM\footnote{\url{https://hedam.lam.fr}},][]{Shirley21}, where the field is referred to as LOCKMAN-SWIRE. The image obtained after cropping the irregular edges and removing the $\sim 40$ galaxies whose peak brightness is over 6 MJy/sr is called $d_L$ and is presented in Fig.~\ref{fig:Herschel_observations}. The LH image is also a square with sides of 1144 pixels corresponding to 1.91 degrees on the sky. The pixel size is $6^{\prime\prime}$ as for the Spider maps.

\subsection{Power spectra}
\label{subsec:power_spec}

\begin{figure}[h!]
    \centering
    \includegraphics[width=\hsize]{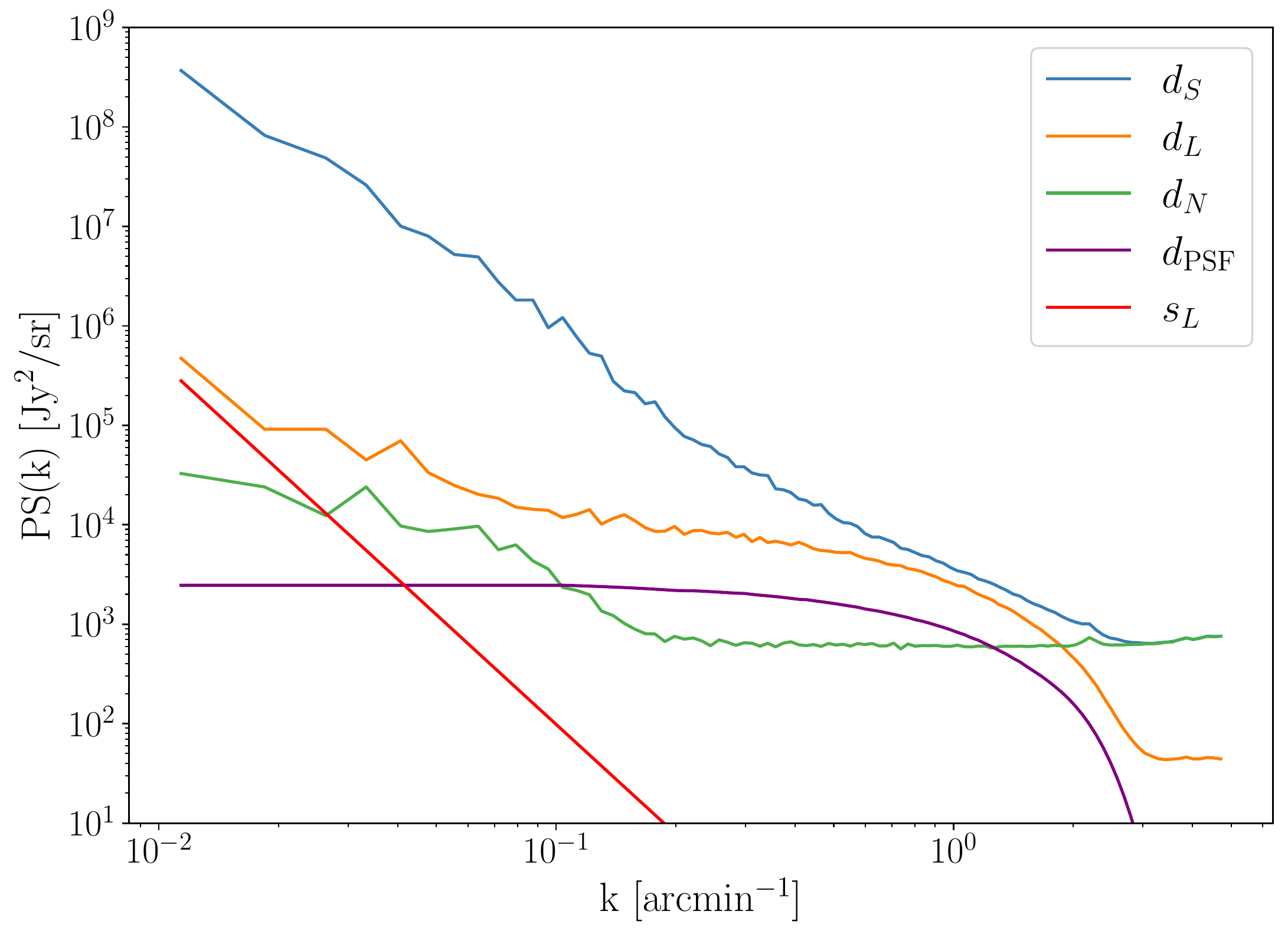}
    \caption{Power spectra of {\it Herschel} SPIRE maps. The power spectra of the Spider map ($d_S$) and data noise ($d_N$) are compared to those of the LH field ($d_L$) with the dust contribution, $s_L$, as estimated by \citet{Viero13}. The spectrum of an observation of Neptune ($d_{\mathrm{PSF}}$), plotted with an arbitrary normalization, indicates the beam attenuation at high $k$. The wavenumber in abscissa is $k = 1 / \theta$. These spectra show that the CIB and noise components dominate the dust signal at the smallest scales in the $d_S$ map.}
    \label{fig:PS_Herschel}
\end{figure}

Figure~\ref{fig:PS_Herschel} presents the power spectra of the $d_S$, $d_L$, and $d_N$ maps. The figure also includes the power spectrum of a SPIRE observation of Neptune at 250\micron. As in \citet{Mamd10}, we consider Neptune as a point-like source and use this observation to compute the power spectrum of the point spread function (PSF).  

The noise spectrum is flat for $k> 0.2\, \mathrm{arcmin}^{-1}$.  
The flattening of the $d_S$ and $d_L$ spectra to different constant values at high $k$ indicates that the noise power is one order of magnitude larger for the Spider observation. 
The $d_L$ spectrum shows the progressive attenuation of the sky emission by the telescope beam for $k > 1 \, {\rm arcmin}^{-1}$. This attenuation is not apparent for the $d_S$ map because the noise amplitude is higher. 

We checked that the power spectrum for the LH field is consistent with the one presented by \citet{Viero13}.
In Fig.~\ref{fig:PS_Herschel}, we also show the power spectrum  of the dust emission in the LH field as estimated by \citet{Viero13}. We used their Eqs.~(8) with $P_0 = 4.5 \times 10^5 Jy^2/sr^{-1}$ and $\alpha_c = -3.66$. As the dust contribution is small except at the largest scale, the power spectrum of $d_L$ is essentially that of the CIB. The power spectrum of $d_S$ is dust dominated for $k < 0.2\, \mathrm{arcmin}^{-1}$. The CIB contribution is significant for higher $k$. 

\section{Component separation method}
\label{sec:comp_sep}

In this section, we explain the procedure that allows us to derive a statistical model of the dust emission from the Spider \textit{Herschel} observations. We present the mathematical formalism of the problem in Sect.~\ref{subsec:maths} and our algorithm implementation in Sect.~\ref{subsec:algorithm}.

\subsection{Mathematical formalism}
\label{subsec:maths}

We want to separate the dust emission from the CIB and the \textit{Herschel} instrumental noise. This problem can be written as a system of three equations involving the three observational maps: $d_S$, $d_L$, and $d_N$. The sky maps $d_S$ and $d_L$ include the dust, CIB, and noise, while the difference map $d_N$ is a noise map that is free from dust and from the CIB. We thus write:
\begin{equation}
    \begin{cases}
    &d_S = s_S + c_S + n_S, \\
    &d_L = s_L + c_L + n_L, \\
    &d_N = n_N, 
\end{cases}
\label{eq:maps}
\end{equation}
where $s$, $c,$ and $n$ terms represent the dust, CIB, and noise contributions to each image, respectively. 

Our scientific objective is twofold. First, we aim to derive a statistical model of the dust emission from the Spider observation expressed in terms of WPH statistics. Second, we aim to derive an estimate of the Spider dust map $s_S$ as one specific realization of the dust model. 

We simplify Eq.~\ref{eq:maps}, neglecting $s_L$ with respect to $c_L$ and $s_S$, and $n_L$ with respect to $n_S$ and $c_L$. These simplifications are supported by the comparison of the $d_S$, $d_L$, and $s_L$ spectra in Fig.~\ref{fig:PS_Herschel}. In the following, we also consider that the CIB is a statistically isotropic signal on the celestial sphere, ignoring its correlation with the large-scale structure of the universe \citep{Planck_XVIII_2013,Serra14}.

We rewrite Eq.~\ref{eq:maps} as: 
\begin{equation}
    \begin{cases}
        \label{eq_maps_reduced}
        &d_S = s_S + c_S + n_S,  \\
        &c^\prime_{L} \equiv d_L+d_N = c_L + n_N, 
    \end{cases}
\end{equation}
where we have introduced the map, $c^\prime_{L}$, which represents a hypothetical observation of LH with the data noise statistically matching that of the Spider observation. We also introduce $c^\prime_S = c_S + n_S$ to rewrite Eq.~\ref{eq_maps_reduced} in terms of the WPH statistics:
\begin{equation}
    \begin{cases}
        \label{eq_WPH}
        &\Phi(d_S) = \Phi(s_S + c^\prime_S), \\
        &\Phi(c^\prime_{S}) = \Phi(c^\prime_L), 
    \end{cases}
\end{equation}
where $\Phi(m)$ are the WPH statistics (see Appendix~\ref{app:WPH_statistics}) computed for a map, $m$. The second equation follows from the isotropy of the CIB on the celestial sphere.

These equations lead us to consider two statistical processes: $S$ associated with the dust emission and $C^\prime$ associated with the sum of the CIB and the Spider data noise. Hereafter, we refer to $S$ and $C'$ as the dust and contamination, respectively. The map $s_S$ is a realization of $S$, while the maps $c^\prime_S $ and $c^\prime_L $ are distinct realizations of $C'$. 
From the map, $c^\prime_{L}$, we can directly compute the WPH statistics that characterize $C^\prime$. To characterize $S$ we use a statistical component separation algorithm explained below.

\subsection{Algorithm principle}
\label{subsec:algorithm}

The framework we used for statistical component separation has already been studied through the denoising of \textit{Planck} interstellar dust polarization maps on a flat sky using WPH \citep{regaldo-saint_blancard_new_2021} and on the sphere, using an implementation of cross-WST on healpix \citep{delouis_non-gaussian_2022}. We followed and extended the work of \citet{regaldo-saint_blancard_new_2021}. This method consists of an iterative minimization in pixel space of a loss function. We performed an iteration on a dust map, $u,$ to converge to an estimator $\Tilde{s_S}$ of $s_S$, such that the $u+c^\prime_S$ map and the $d_S$ map become “close enough” in terms of WPH statistics.

The quantitative description of the algorithm is described by the loss function. The expression of our loss function $\mathcal{L}(u)$ is:
\begin{equation}
    \displaystyle \mathcal{L}(u) = \mathcal{L}_1(u) + \alpha \ \mathcal{L}_2(u),
\end{equation}
where 
\begin{eqnarray}
\begin{aligned}
\label{eq_loss_terms}
&\displaystyle \mathcal{L}_1(u) = \langle || \Phi(u+c^\prime_{S,i}) - \Phi(d_S) ||^2 \rangle_i, \\
&\displaystyle \mathcal{L}_2(u) = || \Phi(d_S-u) - \Phi(c^\prime_S) ||^2,
\end{aligned}
\end{eqnarray}
where $||\cdot||$ stands for the Euclidean norm and $\langle \cdot \rangle_i$ stands for the average over $i$ and $\alpha$ is a real number that will be chosen to balance the two loss terms. The set $\displaystyle \{ c^\prime_{S,i} \}_{i<N}$ represents $N$ sky maps generated from the WPH statistics $\Phi(c^\prime_{L})$ (see Appendix~\ref{app:generative}). We use the WPH statistics defined as in \citet{regaldo-saint_blancard_generative_2022} and refer to Appendix~\ref{app:WPH_definition} for technical details. The WPH statistics used in this paper are made of 3441 coefficients. This definition includes a normalization which is done by using two reference maps, $m_1$ and $m_2$, one for each of the loss terms. 

Our loss function, $\mathcal{L}(u),$ is composed of two non-orthogonal terms. The $\mathcal{L}_1$ term is the loss function used in \citet{regaldo-saint_blancard_new_2021}. It follows directly from the top line of Eq.~(\ref{eq_WPH}). It constrains the $u$ map such that if we add an independent realization of $c^\prime_S$, the sum has the same WPH statistics as $d_S$. We averaged over several syntheses of $c^\prime_S$ so that the separation does not depend on the deterministic properties of one specific realization of $C'$. The $\mathcal{L}_2$ term follows from the bottom line of Eq.~(\ref{eq_WPH}). It constrains the $d_S-u$ map, which we want to converge to $c^\prime_S = d_S - s_S$ to have the same WPH statistics as $c^\prime_S$. In \citet{regaldo-saint_blancard_new_2021}, the contamination was a piece-wise Gaussian noise, while in our case, $C^\prime$ is a non-Gaussian process. Experimentally, we found that the $\mathcal{L}_2(u)$ term is necessary to account for the non-Gaussianity of $C^\prime$.

Recently, \citet{regaldo-saint_blancard_generative_2022} introduced cross-WPH statistics to capture the non-Gaussian correlations between different maps. They applied it successfully to the building of multifrequency dust generative models. Simultaneously, \citet{delouis_non-gaussian_2022} developed cross-WST statistics on the sphere and applied it to the denoising of \textit{Planck} interstellar dust polarization full-sky maps. The success of this method lies in the use of cross-statistics between half-missions maps with distinct data noises, and the TE dust correlation \citep{Planck_XXX}, which makes use of the high signal-to-noise ratio (S/N) of the dust total intensity map. In our case, the same CIB sky signal is present in both half-missions maps, but we could have added a loss term based on the dust-\HI correlation. We did not implement such a term because the \HI data \citep{blagrave_dhigls_2017} has a lower angular resolution than the \textit{Herschel} maps and a higher noise level.

To implement the algorithm, we need to chose $\alpha$, the two reference maps, $m_1$ and $m_2$, used for the loss normalization and the initial value, $u_0$, of the $u$ map. Theses choices will be discussed in the following sections. At the end of the optimization, we converge to a map $\Tilde{s}_S$. The WPH statistics of $\Tilde{s}_S$ constitute the dust statistical model. The model may be used to generate new realizations where the optimal choice of the initial map depends on the scientific objective. 

The maps generated from the statistical model have structure on all scales, including those where the power of the contamination is much larger than that of the dust. Our goal is to reproduce the non-Gaussian dust statistics even on the scales where we do not succeed in reproducing deterministically the true dust map \citep{regaldo-saint_blancard_new_2021, delouis_non-gaussian_2022}. Our approach differs from other attempts to separate the dust and CIB \citep{remazeilles_foreground_2011} in two main ways: our component separation is based on non-Gaussian statistics and we do not seek to minimize the mean squared error in pixel space~\citep[see, e.g.,][]{wiener_extrapolation_1949}.

\section{Validation on mock data}
\label{sec:validation}

To validate our method, we applied our component separation algorithm to mock data. The mock data are introduced in Sect.~\ref{subsec:mock_data}. We compare input and output maps in Sect.~\ref{subsec:separation results}. In Sect.~\ref{subsec:statistical_diagnostic}, we show that the output maps reproduce statistics of the input mock data, which are used as diagnostics in interstellar astrophysics. 

\subsection{Mock data}
\label{subsec:mock_data}

We present how we built the mock data from observations to be as close as possible to the real data. We produce a set of mock maps defined as: 
\begin{equation}
    \label{eq:mock_data}
    d_S^{m,i} = s_S^m + c^{\prime \ m,i}_{S}
,\end{equation}
where $m$ stands for mock data and the $i$ index indicates the different mock maps. Guided by the correlation between dust and gas in the high latitude sky \citep{lenz_large-scale_2019}, we use an integrated intensity map, \IHI\, of 21\,cm data to build $s_S^m$ as a pure dust map free from the CIB. Specifically, we made use of the DF dataset, located at $(\alpha,\,\delta) = (10^{{\mathrm h}}$30$^{{\mathrm m}}$,$73$\degree$48')$ (i.e., centred on the Spider field) that was part of the DHIGLS\footnote{DRAO \HI\ Intermediate Galactic Latitude Survey: \url{https://www.cita.utoronto.ca/DHIGLS/}} \HI\ survey \citep{blagrave_dhigls_2017} with the Synthesis Telescope (ST) at the Dominion Radio Astrophysical Observatory. A detailed description of the DF dataset is presented in Appendix~\ref{app:HI_map}, along with the procedure used to reduce noise in the data using the Gaussian decomposition algorithm {\tt ROHSA} \citep{marchal_rohsa_2019}. The velocity range of integration is -9.7 < v < 26.5 km/s. The \IHI\ map is a square with 312 pixels on each side corresponding to 5.2 degrees on the sky. The sky region covered by the \IHI\ map includes the one covered by the $d_S$ map.

To obtain $s_S^m$, we scaled the \IHI\ map, so that the power spectrum of $d_S^{m,i}$ would approximately match that of $d_S$. The $s_S^m$ map is used as a spatial template of clean dust emission at the \textit{Herschel} resolution. The mock dust map is presented in the top left panel of Fig.~\ref{fig:mock data}. For the contamination, we use the $c^\prime_L$ map that combines two \textit{Herschel} maps: the LH CIB map, $d_L$, and the noise map, $d_N$, of the Spider observation. Since the $s_S^m$ map has only 312x312 pixels, we cut the 1144x1144 contamination map, $c^\prime_L$, into nine independent patches of equal areas of 312x312 pixels. These patches allowed us to take into account the spatial variations of the CIB statistics over the LH region. They were combined into 72 independent pairs; for each pair, we used the first patch to produce the mock data and the second one as input to learn the statistics of the contamination. The mock $d_S^{m,i}$ were thus constructed as the sum of $c^{\prime \ m,i}_{S}$ and $s_S^m$. One pair of maps $\displaystyle \{ c^{\prime \ m,i}_{S},d_S^{m,i} \}$ is plotted next to the $s_S^m$ map in Fig.~\ref{fig:mock data}.

We used the set to run our separation algorithm on these 72 different cases. The mean and the standard deviation of these results allow to verify that the algorithm does not produce a significant bias on the statistics of the dust map. Unfortunately, our validation is based on a small sample because the 72 cases are not independent. However, we are confident that this method, which allows us to take into account the non-Gaussianity of the CIB encoded in the observations, gives a satisfactory check of the significance of any potential bias. Using WPH syntheses (Appendix~\ref{app:generative}), we could have generated more non-Gaussian realizations of the contamination to better sample the chance correlation with the dust mock map, but this possibility was constrained by the required computation time. It is important to take into account the variance of the CIB statistics to verify the algorithm ability to determine the dust statistics when there are small differences between the statistics of the patch, $c^{\prime \ m,j}_{L}$, used to model the contamination, with respect to those of the contamination in $d_S^{m,i}$. This is only indicative because the variance between the contamination patches is not necessarily commensurate with the CIB variance on the sky between the LH and Spider fields.

\subsection{Implementation of the component separation and results}
\label{subsec:separation results}

\begin{figure*}
    \centering
    \resizebox{0.99\hsize}{!}{\includegraphics{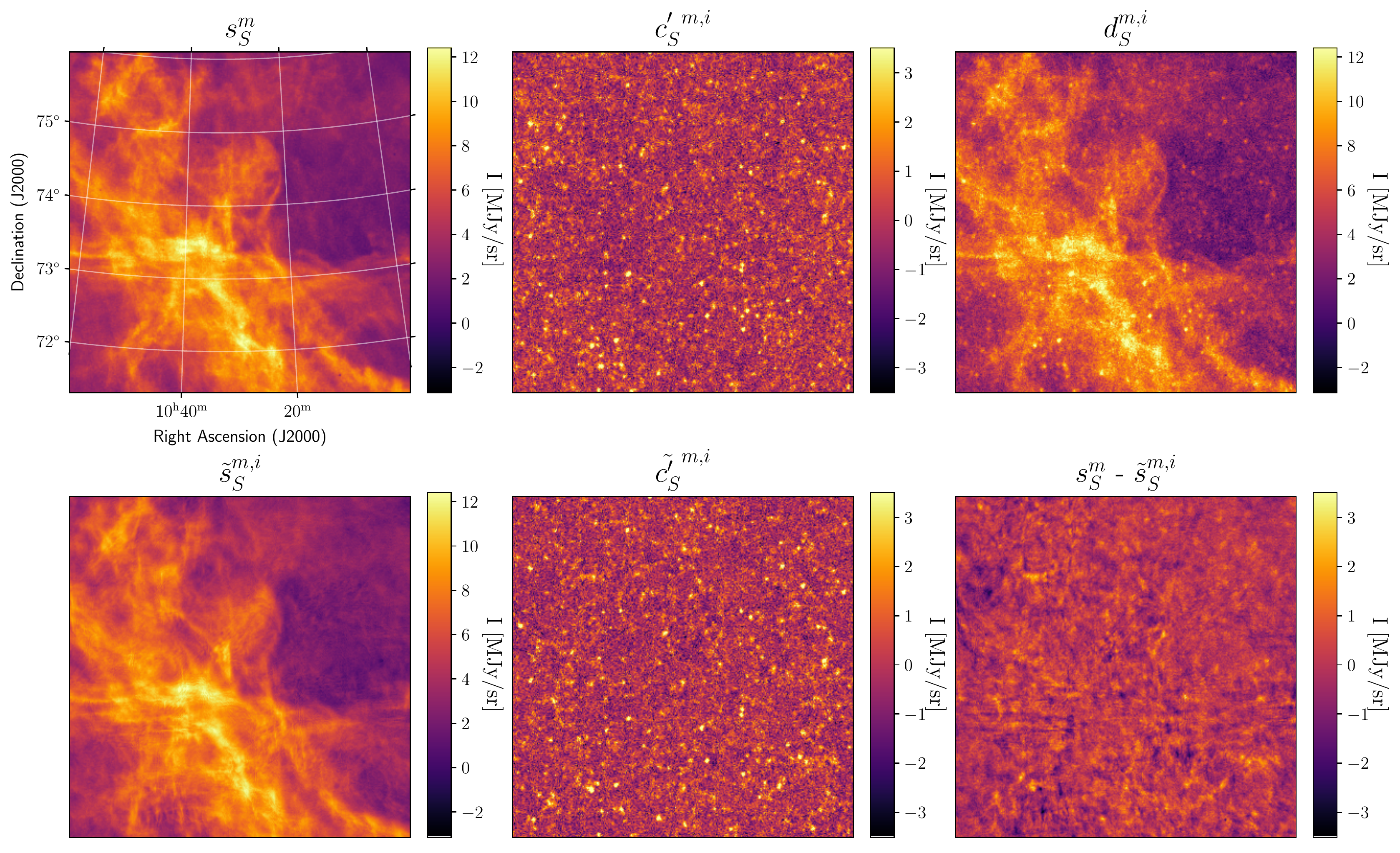}}
    \caption{Input and output maps for the component separation applied to the mock data. The three top images show the input dust map $s_S^m$ (left), one example of the input contamination $c^{\prime \ m,i}_{S}$ (middle), and the mock data $d_S^{m,i}$ (right). The three bottom images show output maps from the component separation: the dust map $\tilde{s}_S^{m,i}$ (left), the contamination $\tilde{c}^{\prime \ m,i}_{S}$ (middle), and the difference $s_S^m - \tilde{s}_S^{m,i}$ (right).}
    \label{fig:mock data}
\end{figure*} 

Here, we describe the implementation of the component separation algorithm before presenting the output images. To apply the algorithm, we made the following choices for the initial map, $u_0$, the maps used for the loss normalizations, $m_1$ and $m_2$, and the inter-loss weight, $\alpha$ (Sect.~\ref{subsec:algorithm}). 

As in \citet{regaldo-saint_blancard_new_2021}, we chose $u_0=d_S^{m,i}$ to obtain an output dust map reproducing the observed map on scales where the dust emission dominates. The $m_1$ and $m_2$ maps are used to normalize the WPH statistics versus scale in $\mathcal{L}_1$ and $\mathcal{L}_2$. The simplest normalization would be to use the mock observation $d_S^{m,i}$ for $m_1$ and a contamination map $c^{\prime \ m,i}_{L}$ for $m_2$. \citet{regaldo-saint_blancard_new_2021}  showed that this choice for $m_1$ is not optimal to reproduce the dust WPH statistics on the smallest scales, where the contamination is dominant. Ideally, we would need to know beforehand the dust power spectrum to give the appropriate weight to the dust WPH statistics. In practice, we ran the algorithm in two steps. First, with $m_1 = d_S^{m,i}$ to converge to a first dust map, $\tilde{s}_{S,0}^{m,i}$, which does not reproduce  the non-Gaussian statistics well, but has the proper dust power spectrum. For this first run, we only used the first loss term, $\mathcal{L}_1$, computed on a subset of WPH statistics that only contain power spectrum-like terms.

\begin{figure}[!ht]
    \centering
    \includegraphics[width=\hsize]{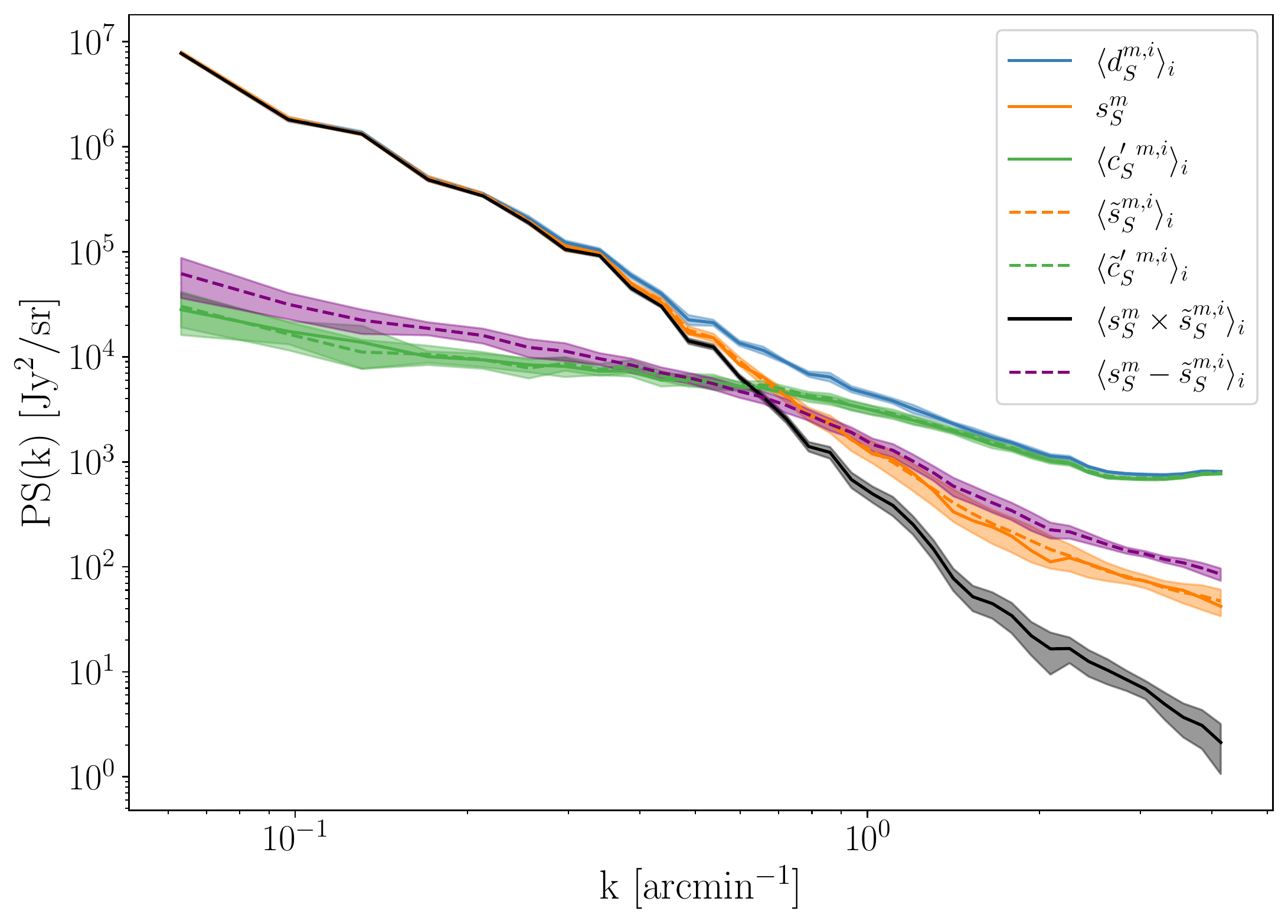}
    \caption{Power spectra of the input and output maps of 
    the component separation applied to the mock data. The figure shows the power spectra of $d_S^{m,i}$, $s_S^m$, $c^{\prime \ m,i}_{S}$, $\tilde{s}_S^{m,i}$, $\tilde{c}_S^{\prime \ m,i}$, and $s_S^m - \tilde{s}_S^{m,i}$ as well as the cross spectrum between $s_S^m$ and $\tilde{s}_S^{m,i}$. The notation $\langle \cdot \rangle_i$ represents the mean of the spectra computed over the 72 separation runs. The colored bands represent $\pm 1\sigma$ error-bars, computed as the standard deviation of these spectra. The component separation allows to recover the power spectra of the input maps within statistical uncertainties.}
    \label{fig:PS_validation}
\end{figure}

Second, we ran the algorithm with the two loss terms with $m_1 = \tilde{s}_{S,0}^{m,i}$ and $m_2 = c^{\prime \ m,i}_{L}$. We choose $\alpha$ to give more weight to $\mathcal{L}_1$ than to $\mathcal{L}_2$ at the start of the gradient descent because we want the first loss to dominate at this point. This is because the second loss term is initially applied to an identically zero map, since $u_0=d_S$, and thus contains no information. We expect $\mathcal{L}_2$ to help $d_S - u$ to converge to the statistics of the contamination in a second stage. We set $\alpha$ such that $\displaystyle \mathcal{L}_1(u_0) \sim 10 \alpha \ \mathcal{L}_2(u_0)$ and find that the result of the separation is only weakly dependent on this specific choice. For the validation algorithm, a run takes about 2 hours on a 32 GB GPU.

The WPH statistics of $d_S^{m,i}$, $s_S^m$, $c^{\prime \ m,i}_{S}$, $\tilde{s}_S^{m,i}$, and $\tilde{c}_S^{\prime \ m,i}$ are presented in Fig.~\ref{fig:WPH_validation} and discussed in Appendix~\ref{app:WPH_validation}. The figure shows that our specific choice of $\{ u_0,m_1,m_2,\alpha \}$ allows us to 
\begin{figure}[h!]
    \centering
    \includegraphics[width=\hsize]{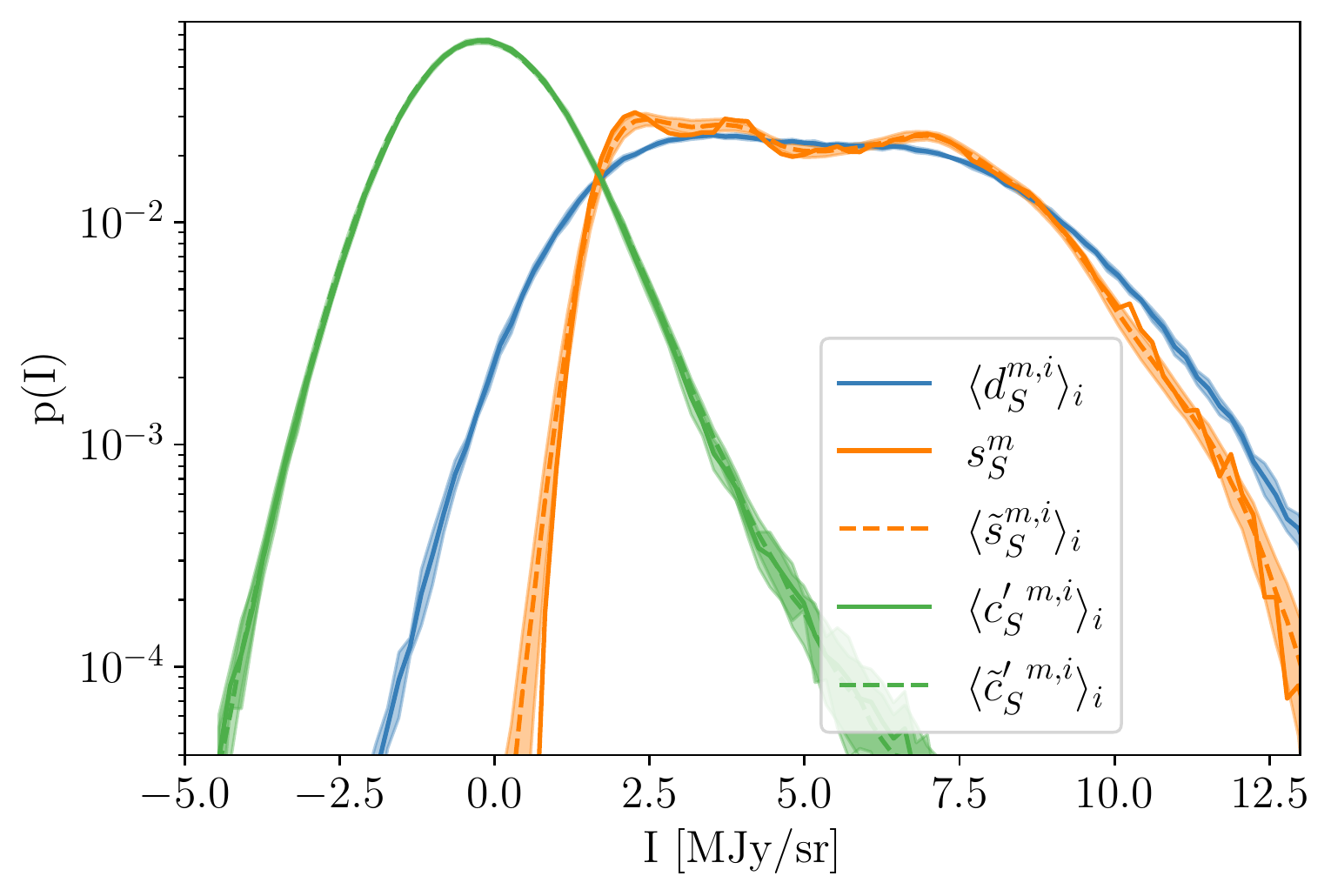}
    \caption{PDFs of the dust intensity for the input and output maps of the component separation applied to the mock data. The PDFs of $d_S^{m,i}$, $s_S^m$, $\tilde{s}_S^{m,i}$, $c^{\prime \ m,i}_{S}$, and $\tilde{c}_S^{\prime \ m,i}$ are compared. The color bands represent $\pm 1\sigma$ error-bars.}
    \label{fig:PDF_validation}
\end{figure}
recover the WPH statistics of dust within error bars at all scales. This validates our algorithm with respect to our primary goal, which is to properly recover the non-Gaussian statistics of dust emission.
\begin{figure*}
    \centering
    \resizebox{0.99\hsize}{!}{\includegraphics{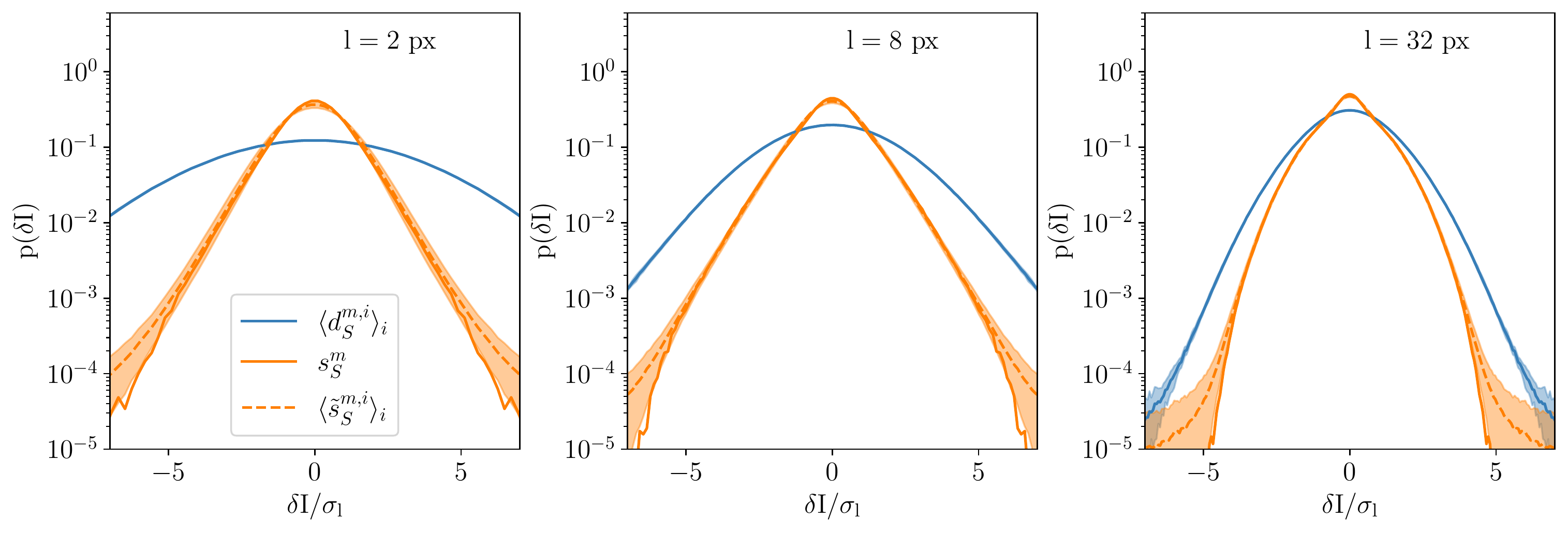}}
    \caption{PDFs of increments of the dust intensity for the input and output maps of the component separation applied to the mock data. The PDFs of $d_S^{m,i}$, $s_S^m$ and $\tilde{s}_S^{m,i}$ are compared for three lag values: 2 pixels (left), 8 (middle), and 32 (right). The PDFs are plotted as a function of $\delta I / \sigma_l$, where $\sigma_l$ is the standard deviation of the increments of $s_S^m$ at lag $l$. The colored bands represent $\pm 1\sigma$ error-bars.}
    \label{fig:increment_PDF_validation}
\end{figure*} 
\begin{figure*}
    \centering
    \resizebox{0.99\hsize}{!}{\includegraphics{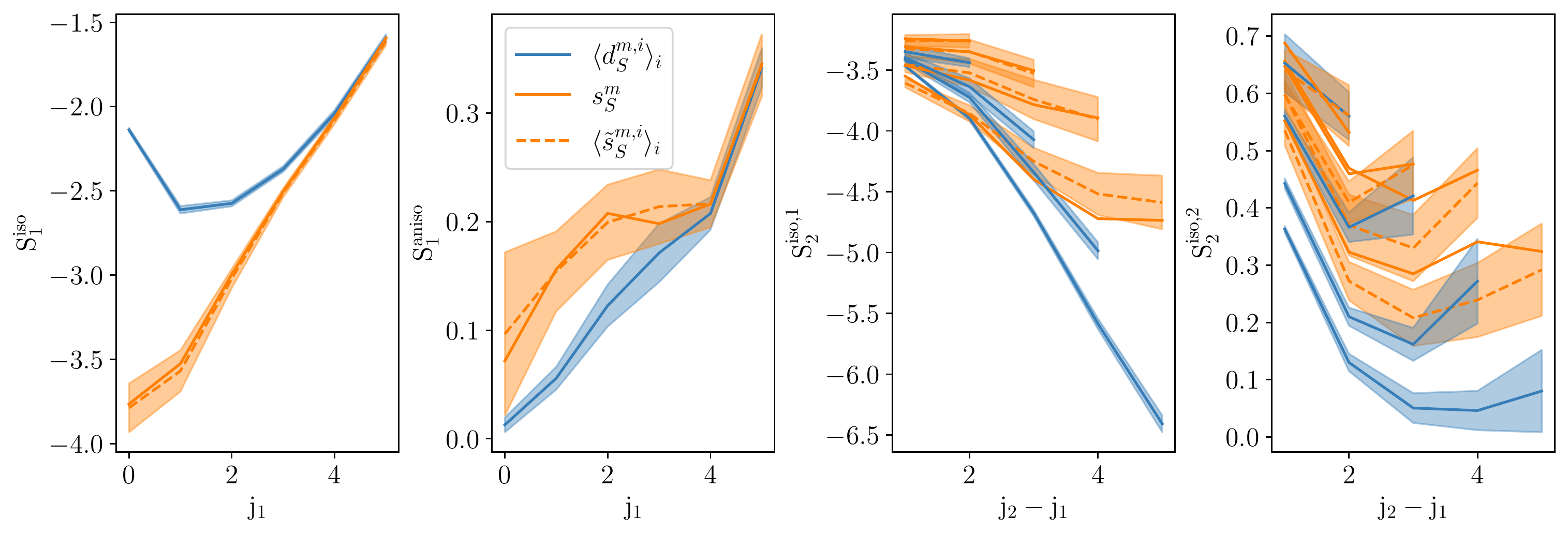}}
    \caption{RWST statistics of the input and output maps of the component separation applied to the mock data. The RWST statistics of 
    $d_S^{m,i}$, $s_S^m$, and $\tilde{s}_S^{m,i}$ are compared. From left to right, the figure panels display $S_1^{iso}$ (power at a given dyadic scale), $S_1^{aniso}$ (level of anisotropy at a given dyadic scale), $S_2^{iso,1}$ (couplings between different dyadic scales), and $S_2^{iso,2}$ (angular modulation of the couplings between different dyadic scales). The variable $j$ corresponds to the physical scale of $2^{j}$. The $S_2^{iso,1}$ and $S_2^{iso,2}$ are normalized with respect to the $S_1^{iso}$ and are plotted as a function of the scale ratio $j_2-j_1$ for $j_1 \in [0,J-1]$ and $j_2 \in [j_1 + 1,J-1]$.  Each curve corresponds to a given $j_1$. The colored bands represent the $\pm 1\sigma$ error-bars.}
    \label{fig:RWST_validation}
\end{figure*} 
Figure~\ref{fig:mock data} presents the output maps of our separation algorithm applied on $d_S^{m,i}$.
Comparing by eye $\tilde{s}_S^{m,i}$ with $s_S^m$, we see that we have efficiently removed the contamination without losing the non-Gaussian structure of the dust emission down to the smallest scales. The comparison of $\tilde{c}^{\prime \ m,i}_{S}$ and $c^{\prime \ m,i}_{S}$ is equally satisfactory. The standard deviation of $s_S^m$, $c^{\prime \ m,i}_{S}$, $d_S^{m,i}$, $\tilde{s}_S^{m,i}$, $\tilde{c}^{\prime \ m,i}_{S}$, and $s_S^{m,i} - \tilde{s}_S^{m,i}$ are  2.39, 1.17, 2.61, 2.34, 1.17, and 0.64, respectively. The $d_S^{m,i}$ mock map does not exactly resemble  the \textit{Herschel} $d_S$ one because the first covers a much larger sky area than the second. The sky region covered by the $d_S$ map corresponds approximately to the quarter at the bottom left of $d_S^{m,i}$.

In Fig.~\ref{fig:PS_validation}, we present the power spectra of the six maps shown in Fig.~\ref{fig:mock data}, plus the cross spectrum between $s_S^m$ and $\tilde{s}_S^{m,i}$. These spectra have been computed on apodized maps and binned in order to lower the statistical variance at large values of $k$. Given the power spectra of $s_S^m$ and $\tilde{s}_S^{m,i}$ are very close at all scales, the separation is able to recover the power spectrum of the dust map more than one order of magnitude under the contamination. The power spectra of $c^{\prime \ m,i}_{S}$ and $\tilde{c}^{\prime \ m,i}_{S}$ are also very close, which could be expected since its directly constrained by the $\mathcal{L}_2$ loss term. It shows the ability of our method to extract a realistic contamination from the components mixture. 

The cross spectrum $\langle s_S^m \times \tilde{s}_S^{m,i} \rangle_i$ shows the transition from a deterministic to a statistical separation at the scale of $k = 0.7 \ \text{arcmin}^{-1}$, where the dust power becomes lower than that of the contamination. This transition is similar to that reported by \citet{regaldo-saint_blancard_new_2021} and \citet{delouis_non-gaussian_2022} for the denoising of \textit{Planck} dust polarization maps.  
On the smallest scales, $s_S^m$ and $\tilde{s}_S^{m,i}$ are two independent realizations of the dust statistics. This explains the factor of 2 difference between the power spectrum of $s_S^m - \tilde{s}_S^{m,i}$ and those of $s_S^m$ and $\tilde{s}_S^{m,i}$. Indeed, we see coherent features distributed at small scales in the difference map at the bottom right corner of Fig.~\ref{fig:mock data}, which testify to the displacement of structures from $s_S^m$ to $\tilde{s}_S^{m,i}$.

\begin{figure*}
    \centering
    \resizebox{\hsize}{!}{\includegraphics{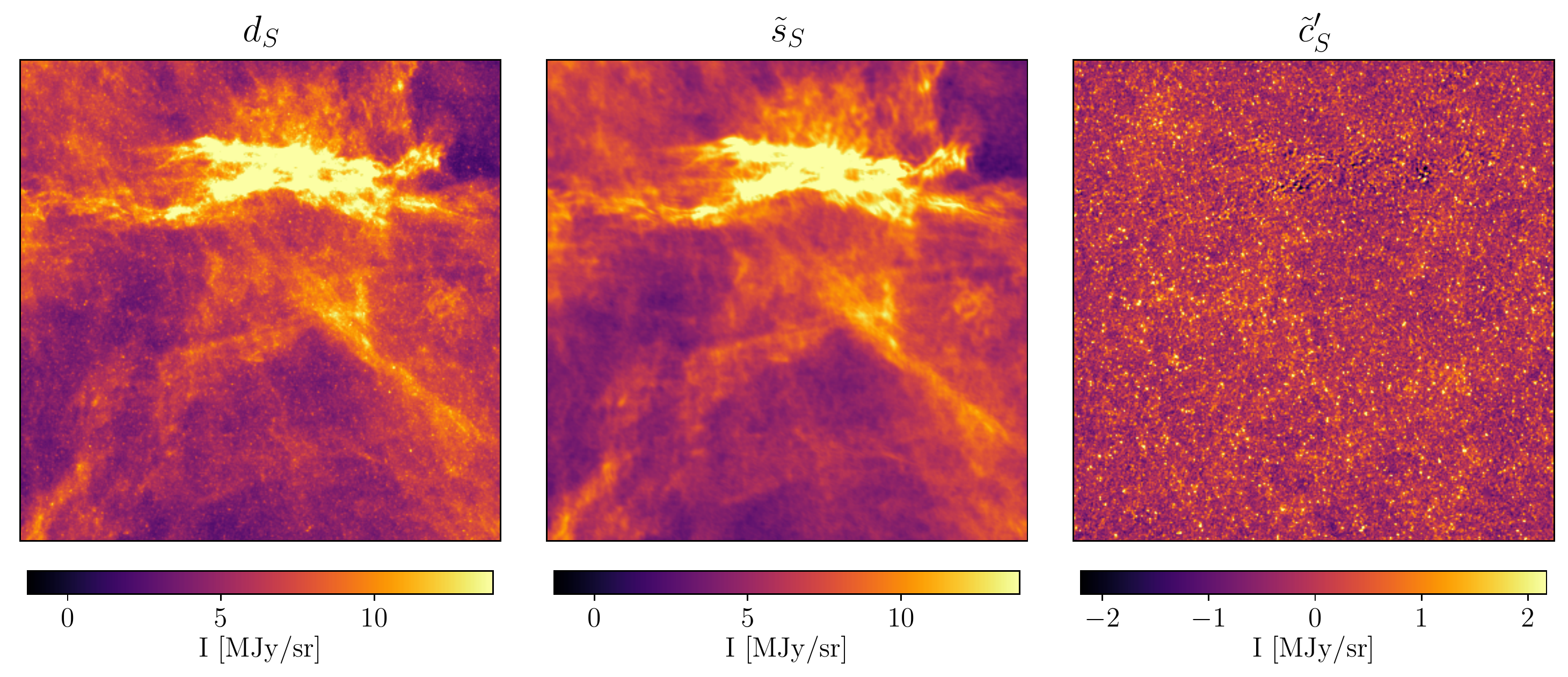}}
    \caption{Input and output maps for the component separation applied to \textit{Herschel} SPIRE maps at 250\micron. Left: Map of the Spider field $d_S$. Center: Output dust map $\Tilde{s}_S$. Right: Output contamination map $\tilde{c}_S^{\prime}$. The $d_S$ map is separated into two components: the dust, $\Tilde{s}_S$, and the CIB+noise contamination, $\tilde{c}_S^{\prime}$.}
    \label{fig:images_Herschel_separation}
\end{figure*} 

\subsection{Non-Gaussian diagnostic for dust astrophysics}
\label{subsec:statistical_diagnostic}

We show that the dust output maps reproduce statistics used
as diagnostics to characterize the non-Gaussianity of interstellar imaging observations. 
We considered the probability distribution functions (PDFs) of the intensity and its spatial increments and the reduced wavelet scattering transform \citep[RWST,][]{allys_rwst_2019}. The notation $\langle \cdot \rangle_i$ represents the mean of the statistics computed over our set of 72 separations (Sect.~\ref{subsec:mock_data}), whereas the error bar is the standard deviation.

Figure~\ref{fig:PDF_validation} presents the PDFs of the dust intensity, which is commonly used as a diagnostic of the  structure of molecular clouds \citep{burkhart_diagnosing_2021,lombardi_molecular_2015}. The PDF of $d_S^{m,i}$ is clearly biased towards low and high values by the contamination. After the component separation we recover very well, within statistical uncertainties, the PDF of the dust map $s_S^m$. The PDF of the contamination map $c^{\prime \ m,i}_{S}$ is also well recovered.

Figure~\ref{fig:increment_PDF_validation} shows PDFs of the increments of the dust emission for three lags. The increment at a position $\Vec{x}$ for a lag $l$ in pixels is the set of differences $\delta I_{\Vec{l}}(\Vec{x}) = I(\Vec{x}) - I(\Vec{x}+\Vec{l})$ with $l<|\Vec{l}|<l+1$. Such PDFs have been computed on velocity maps \citep{hily-blant_dissipative_2008} and polarization maps \citep{regaldo-saint_blancard_new_2021} to characterize the intermittence of turbulence in the ISM. The PDF of $d_S^{m,i}$ is far more Gaussian than that of $s_S^m$, especially for small pixel lags where the impact of the contamination is the largest. For the three lags, the PDF of $\tilde{s}_S^{m,i}$ matches the one of $s_S^m$ within statistical uncertainties over at least three orders of magnitude.

The RWST statistics are low-dimensional non-Gaussian interpretable statistics that has found many applications in astrophysics \citep{allys_rwst_2019,regaldo-saint_blancard_statistical_2020,Saydjari21}. They are briefly presented in Appendix~\ref{app:RWST}. We use here the $S_1^{iso}$, $S_1^{aniso}$, $S_2^{iso,1}$, and $S_2^{iso,2}$ coefficients. Figure~\ref{fig:RWST_validation} presents those RWST statistics for $d_S^{m,i}$, $s_S^m$ and $\tilde{s}_S^{m,i}$.The $S_1^{\text{iso}}$ and $S_1^{\text{aniso}}$ characterize  the amplitude as well as the level of anisotropy as a function of scales, respectively. The $S_2^{\text{iso,1}}$ characterize the couplings between different scales, while the $S_2^{\text{iso,2}}$ describe their isotropic angular modulation \citep{allys_rwst_2019}. 
The RWST statistics of $d_S^{m,i}$ are strongly biased by the contamination. We can also notice that the RWST statistics of $s_S^m$ and $\tilde{s}_S^{m,i}$ are very close with respect to that of $d_S^{m,i}$ at most of the scales, but we can notice small discrepancies at the smallest scales. 

In the three examples presented in Figs.~\ref{fig:PDF_validation}, \ref{fig:increment_PDF_validation}, and \ref{fig:RWST_validation} we see that these non-Gaussian statistics of $s_S^m$ are very well recovered through our component separation method. This result 
illustrates the possibility of using the WPH model to determine non-Gaussian statistics that are not directly constrained in the component separation. It demonstrates the relevance of the WPH statistics for astrophysics. We point out the large difference between the statistics of $d_S^{m,i}$ and those of $s_S^m$, which shows that the component separation step is essential to build a precise non-Gaussian statistical model of the dust emission.   

\section{Application to \textit{Herschel} SPIRE observations}
\label{sec:application}

We applied our component separation method to the \textit{Herschel} observation of the Spider field, $d_S$ (Sect.~\ref{subsec:images_Herschel}). The power spectrum of the output map is presented and analyzed in Sect.~\ref{subsec:PS_Herschel_separation}. In Sect.~\ref{subsec:coherent_structures}, we present images that highlight a main non-Gaussian feature of the dust maps: the correlation between structures across scales.

\subsection{Herschel separation results}
\label{subsec:images_Herschel}

Our algorithm is applied to the $d_S$ map as described in Sect.~\ref{subsec:separation results}. We increased the pixel size of the image from 6\arcsec\ to 12\arcsec\ because the dust emission is highly attenuated by the beam on those scales (more than five orders of magnitude at 6\arcsec). This resampling is done using a low-pass filter in Fourier space with $k_{\text{max}} = 2.5 \ \text{arcmin}^{-1}$ equal to the Nyquist frequency for 12\arcsec\ pixels. This filtering does not induce ringing in the map because the beam attenuation on the signal power at $k = 2.5 \ \text{arcmin}^{-1}$ is large (about a factor 65). Furthermore, for the purposes of normalization, it allows us to keep the exact same procedure from the validation on mock data to this case.

The component separation algorithm was applied according to the steps given in Sect.~\ref{subsec:separation results} for the mock data. We choose $u_0 = d_S$ for the initial map. First, we run the algorithm using only the $\mathcal{L}_1$ loss term, computed on a subset of WPH statistics that only contain power spectrum-like terms, with $m_1 = d_S$ to produce an intermediate dust map $\Tilde{s}_{S,0}$. Second, we run the algorithm using the two loss terms with $m_1 = \Tilde{s}_{S,0}$ and $m_2 = c^{\prime}_L$. We set $\alpha$ such that $\displaystyle \mathcal{L}_1(u_0) \sim 10 \alpha \ \mathcal{L}_2(u_0)$. In this case, a run of the algorithm takes about 8 hours on a 32 GB GPU.

In Fig.~\ref{fig:images_Herschel_separation}, we present the output maps of our separation algorithm applied on the Spider \textit{Herschel} observation. From a visual point of view, the results are very satisfactory: it indeed seems that most of the CIB contamination has been identified as such. It is notably the case of all galaxies that are individually resolved, which is not surprising because they have an extremely clear non-Gaussian signature, both in terms of frequency and spatial location. Conversely, it does not seem that structures related to Galactic dust have leaked to the reconstructed contamination above the scales where it begins to dominate. This is impressive for results that have been obtained from three observational patches only. It should be noted, however, that the CIB reconstruction has visual defects at the position of the very bright horizontal Galactic dust structure at the top of the image. We believe that this is due to the limitation of considering the dust emission as a statistically homogeneous signal\footnote{A solution to this problem is to impose multiple constraints during the component separations by using different local masks, as done in~\cite{delouis_non-gaussian_2022}. However, we did not implement this approach due to the lack of a sufficient number of pixels}. 

The WPH statistics of the output and input Spider maps $\Tilde{s}_S$ and $d_S$ are discussed in Appendix~\ref{app:WPH_validation} and compared in Fig.~\ref{fig:WPH_Herschel_separation}. The WPH statistics of $\Tilde{s}_S$ constitute our statistical model of the dust emission. Figure~\ref{fig:WPH_Herschel_separation} also presents the statistics of $c^{\prime}_{L}$ and $\tilde{c}_S^{\prime}$. The main scientific results of the component separation algorithm are the statistics of the dust emission and, in particular, its WPH statistics. They are obtained from the $\Tilde{s}_S$ map, which is a realization of the estimated WPH statistical model of the dust emission, conditioned by its large scales, where the dust power is dominant. The $\Tilde{s}_S$ map is then highly correlated at large scales with the $d_S$ map, up to a scale where there is a transition from a deterministic to a statistical behavior \citep{regaldo-saint_blancard_new_2021, delouis_non-gaussian_2022}. 

From the WPH statistical model, we can generate diverse dust maps depending on the specific choice of the initial map, $u_0$, and the scientific goal. For example, we can use an \HI\ map as initial condition to ensure that it is decorrelated from the CIB. For other statistical applications, it is useful to have multiple realizations of the dust emission. In terms of this ability to generate various dust maps, our method differs from deterministic component separations.

Once we ran the component separation algorithm on the \textit{Herschel} observation, we obtained a dust map that provides us with an estimate of the non-Gaussian statistics of the dust emission, which are not biased by the CIB and noise. We then estimated the error bars on these dust statistics as follows. We did not use the error bars determined on the mock data because the statistics of the mock dust map only approximate the true statistics of the dust emission, in particular, because it is based on an \HI\ observation with a lower resolution than the one of the \textit{Herschel} observations. First, we used the WPH synthesis method (described in Appendix~\ref{app:generative}) to synthesize ten new realizations of the dust from our separated dust map, $\Tilde{s}_S$, and ten new realizations of the contamination from the contamination map $c^{\prime}_{L}$. Summing these new dust and contamination maps, we obtain ten mock mixture maps, to which we apply the component separation algorithm to obtain ten new separated dust maps. The standard deviations of the statistics computed over these ten dust maps are then taken as error bars. We note that in this procedure, we cannot use patches of the LH and noise maps to define different contamination samples because the Spider image is as large as them. Unfortunately, this does not allow us to take into account the spatial variations of the contamination statistics in our estimate of the error bars.

\subsection{Power spectra}
\label{subsec:PS_Herschel_separation}

\begin{figure}[h!]
   \centering
   \includegraphics[width=\hsize]{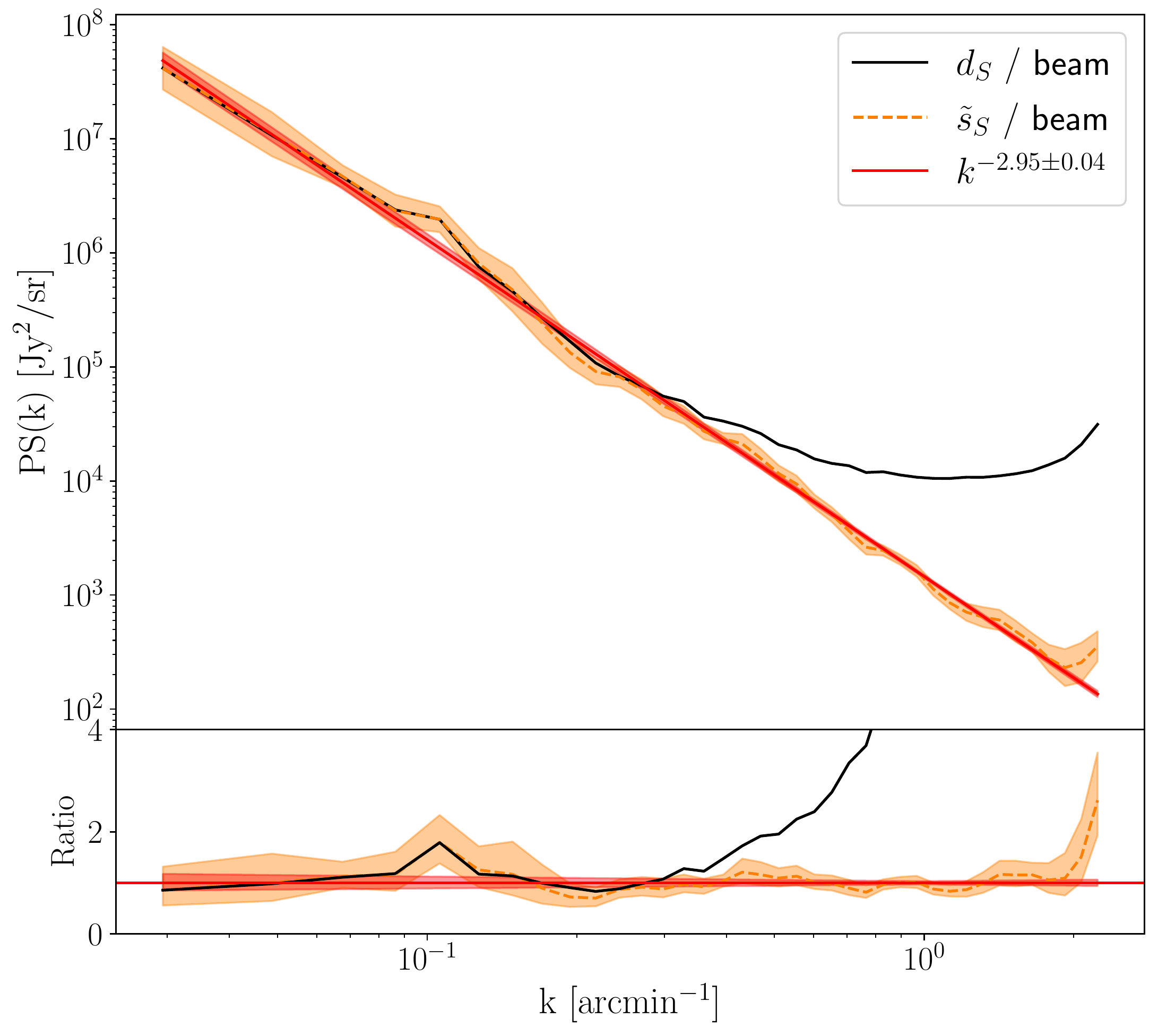}
      \caption{Power spectra of the input and output maps for the component separation applied to \textit{Herschel} SPIRE maps at 250\micron. Top: Beam-corrected power spectrum of $d_S$, beam-corrected power spectrum of $\Tilde{s}_S$ and its power-law fit. Bottom: Ratio with the power-law fit. The colored bands represent $\pm 1\sigma$ error-bars. The component separation extends  the scale range where the dust power spectrum is found to have a power-law shape by a
factor of 6.}
         \label{fig:PS_Herschel_separation}
   \end{figure}

 Figure~\ref{fig:PS_Herschel_separation} presents the beam-corrected power spectrum of $d_S$, the beam-corrected power spectrum of $\tilde{s}_S$, and its power-law fit. These spectra have been computed on apodized maps and binned in order to lower the statistical variance at large $k$. At $k=2 \ \text{arcmin}^{-1}$, the power of $\Tilde{s}_S$ is two orders of magnitude lower than that of $d_S$. 

Figure~\ref{fig:PS_Herschel_separation} shows that the power spectrum of $\tilde{s}_S$ is very close to a power-law after beam correction. To estimate the spectral index of the power-law and its uncertainty, we need to add to the component separation uncertainty the cosmic variance. We must do this because we characterize a statistical process using an observation (one realization of the process) of finite sky area $A$. The standard deviation of the dust power spectrum computed at wavenumber $k$ is:
\begin{equation}
    \sigma_k = \sqrt{\frac{2}{N_k}} P(k),
\end{equation}
where $N_k$ is the number of modes at wavenumber $k$. This leads to 
\begin{equation}
    \sigma_k = \sqrt{\frac{4 \pi}{A}} \sqrt{\frac{1}{4 \pi^2 k \Delta_k}} P(k),
\end{equation}
where the wavenumber $k$ and the bin size $\Delta_k$ are expressed in $\text{rad}^{-1}$ \citep{scott_sample_1994,knox_perrins_1995}.
We compute the total uncertainty as the quadratic sum of $\sigma_k$ and the component separation uncertainty and perform a power-law fit of the beam-corrected power spectrum of $\Tilde{s}_S$ in log scale. 
The bottom panel of Fig.~\ref{fig:PS_Herschel_separation} presents the beam-corrected power spectra of $d_S$ and $\Tilde{s}_S$ divided by the power-law fit\footnote{The peak appearing in the dust power spectra at $k = 0.1 \ \text{arcmin}^{-1}$ corresponds, in Fourier space, to bright spots on an hexagonal pattern. We believe that it reflects the pattern of the SPIRE array of bolometers.}. We can see that the power-law fits well the beam-corrected power spectrum of $\Tilde{s}_S$ until $k=2 \ \text{arcmin}^{-1}$. 
This result is satisfactory because it indicates that the component separation algorithm did not filter out the small scales up to a $k$ value, where the power of $\tilde{s}_S$ is about 2\% of that of $d_S$. The maximum wavenumber up to which the power-law shape is measured is thus increased by almost a decade by the component separation. 

The slope of the spectrum is $-2.95 \pm 0.04$.
We compare this fit result with values measured for the Polaris Flare, a brighter field observed by \textit{Herschel} \citep{Mamd10}, as well as for a diffuse cloud using \textit{Planck}, \textit{WISE} and optical data \citep{miville-deschenes_probing_2016}. In both cases, the CIB is subdominant and is neglected. The power spectrum of the dust maps is a power-law in both fields, but the slope in log scale goes from $-2.65$ in Polaris to $-2.9 \pm 0.1$ for the diffuse cloud. The slope we find is consistent with the latter value.

\subsection{Coherent structures}
\label{subsec:coherent_structures}

The component separation allows us to identify coherent structures that are hidden by the CIB in the SPIRE maps. To illustrate this result, we computed the  maps with increments for different lags, presented in this work. Computed on a map $I$ at a lag $l$, the pixel at the position $\Vec{x}$ of the increment map is the mean of $\displaystyle \{ I(\vec{x}) - I(\Vec{x'}), \ |\Vec{x}-\Vec{x'}|=l \}$. In the following, the dust statistics are computed at 0.4, 0.8, and 1.6 arcmin. These scales corresponds to wavenumbers 2.5, 1.25, and 0.625 $\rm{arcmin}^{-1}$.

\begin{figure*}
    \centering
    \resizebox{0.8\hsize}{!}{\includegraphics{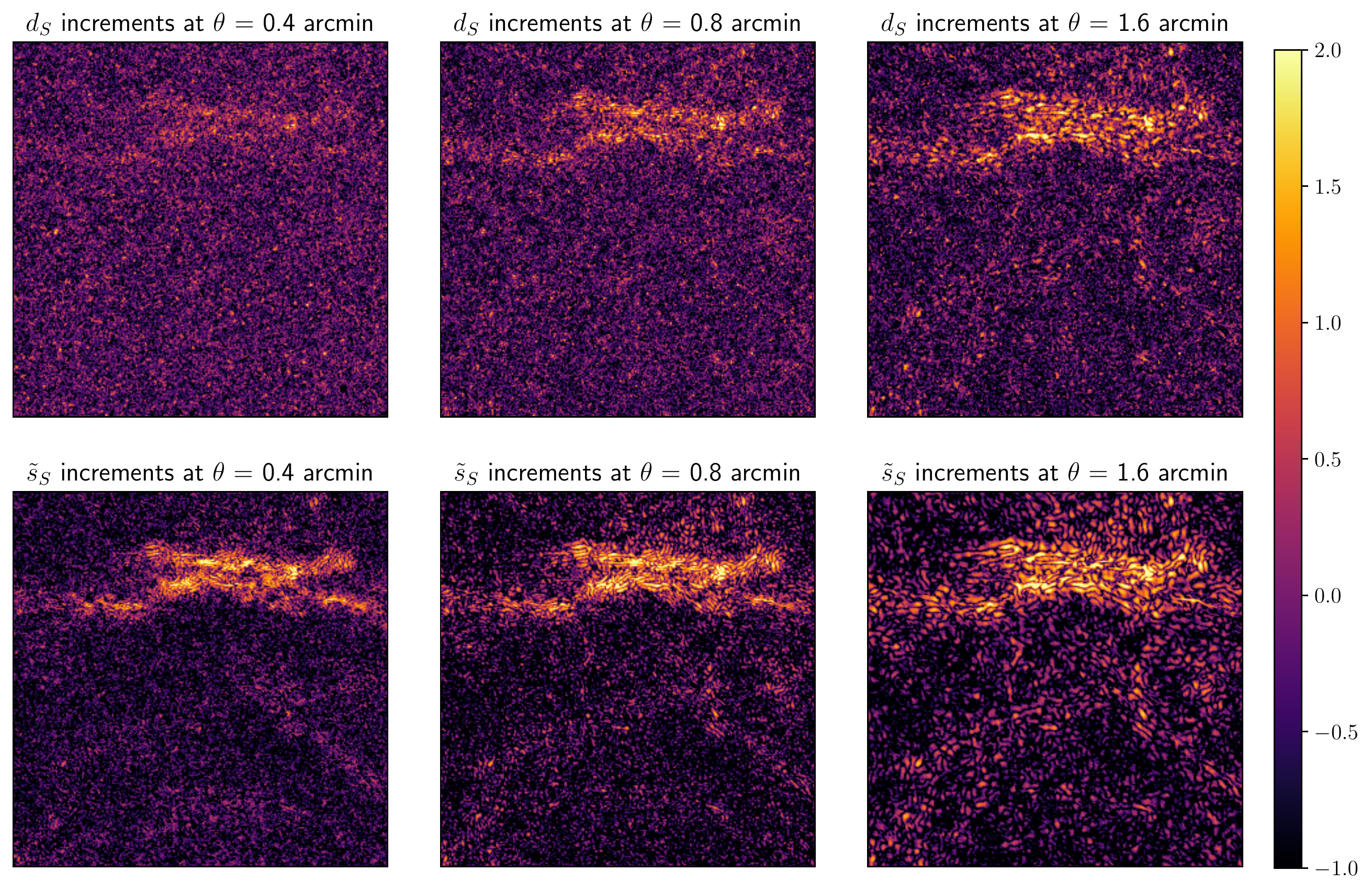}}
    \caption{Maps of increments for the component separation applied to \textit{Herschel} Spider observations. The Neperian logarithm of the absolute value of the increments computed at $\theta =$ 0.4, 0.8, and 1.6 arcmin lags are compared for the input ($d_S$, top row) and output ($\Tilde{s}_S$, bottom row) maps. We substracted the log of the standard deviation of each map. Zooming in on these maps allows for a review of the smallest scales.}
    \label{fig:increment_maps_Herschel_separation}
\end{figure*}

Figure~\ref{fig:increment_maps_Herschel_separation} shows the increment maps of $d_S$ and $\Tilde{s}_S$. One can identify coherent structures on both sets of increments maps, with a much improved contrast for $\tilde{s}_S$. The component separation also highlights the correlation between coherent structures at different scales. This result extends earlier results obtained with wavelet convolution of IRAS sky maps \citep{Abergel96,Jewell01,Mamd07} and with \textit{Herschel} observations on brighter clouds \citep{Robitaille19} to smaller angular scales.

\section{Non-Gaussian statistics of the diffuse dust emission}
\label{sec:non-Gaussian analysis}

Here, we quantify the non-Gaussian statistics of diffuse dust emission in the Spider field  and compare our results with earlier studies.

\begin{figure}[!ht]
   \centering
   \includegraphics[width=\hsize]{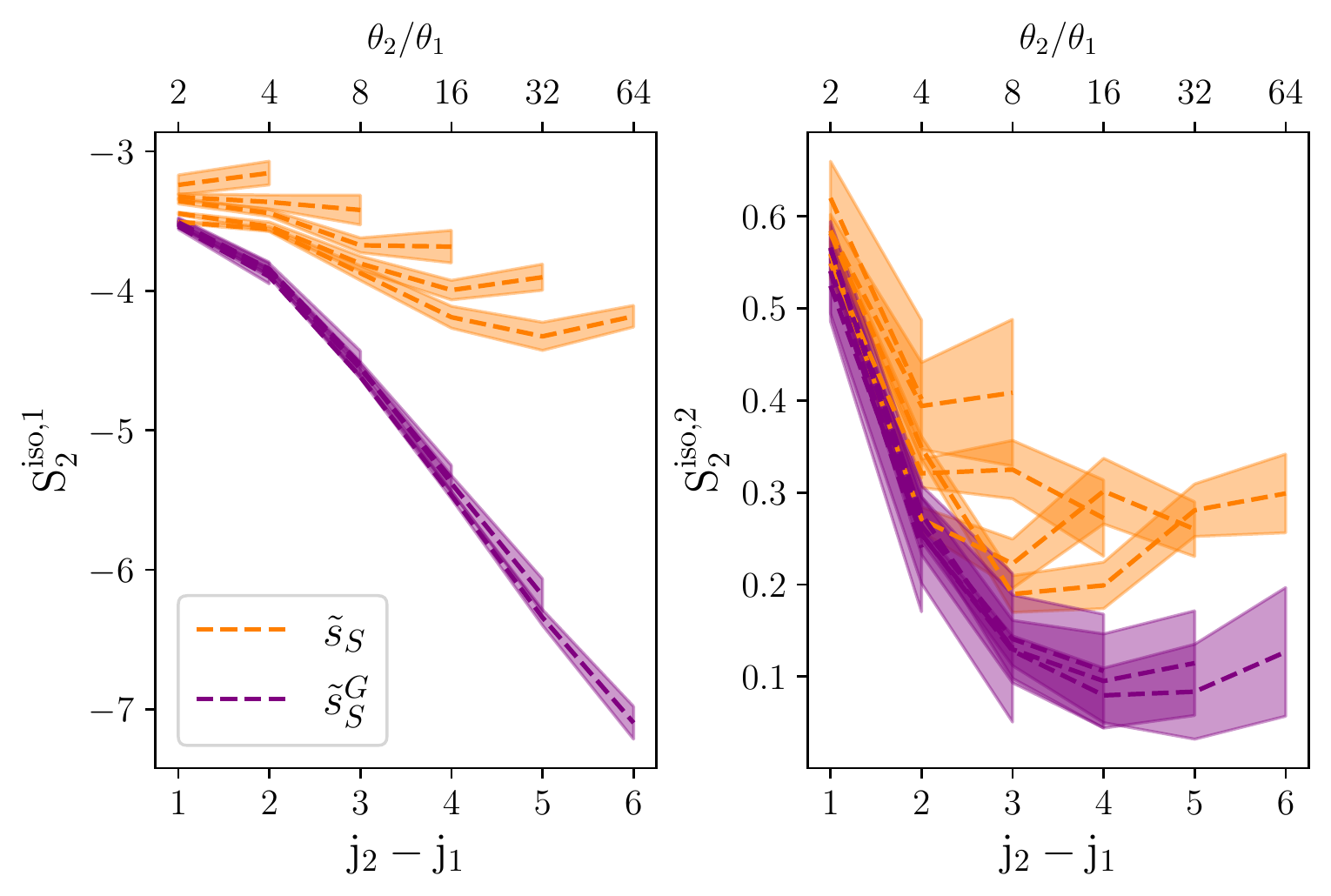}
      \caption{RWST statistics of the output dust map for the component separation applied to \textit{Herschel} SPIRE maps at 250\micron. The $S_2^{iso,1}$ (left) and $S_2^{iso,2}$ (right) coefficients of $\tilde{s}_S$ and $\tilde{s}_S^G$ are compared. These coefficients correspond respectively to the couplings between dyadic scales and their angular modulation. They are normalized with respect to the $S_1^{iso}$ and are plotted as a function of $j_2-j_1$ for $j_1 \in [0,J-1]$ and $j_2 \in [j_1 + 1,J-1]$. Each curve corresponds to a given $j_1$. The $j_2-j_1$ differences on the bottom axes correspond to ratio of angular scales $\theta_2 / \theta_1$ on the top axes. The colored bands represent $\pm 1\sigma$ error-bars. These coefficients testify of the not-scale invariance and the filamentary structure of the output dust map $\tilde{s}_S$.}
         \label{fig:RWST_Herschel_separation}
   \end{figure}
   
\subsection{Coherent structures across scales }
\label{scale_invariance}
   
The RWST statistics (see Appendix~\ref{app:RWST}) provide statistical insights into the multiscale filamentary structure of the cold neutral medium. The dust and CIB separation allows us to expand on the analysis of \textit{Herschel} and \HI data presented by \citet{allys_rwst_2019} and \citet{Lei_Clark22}.

Figure~\ref{fig:RWST_Herschel_separation} presents two RWST coefficients that characterize correlations between scales \citep{allys_rwst_2019}. Both are plotted versus the scale ratio $(j_2-j_1)$. In each plot, the $\Tilde{s}_S$ statistics are compared to those of the map $\Tilde{s}^G_S$ obtained by randomizing the phase of the Fourier transform of $\Tilde{s}_S$. This comparison highlights deviations from Gaussianity. Unlike the Gaussian field, the $S_2^{iso,1}$ of $\Tilde{s}_S$ do not depend solely on $j_2 - j_1$. This result shows that the dust emission is not statistically scale-invariant. The $S_2^{iso,2}$ coefficients quantify the angular modulation of the coupling between scales. The angle dependence relates to the filamentary structure of the diffuse ISM because it measures how the coupling between two scales varies between the aligned and orthogonal orientations. The non-vanishing values of $S_2^{iso,2}$ of $\Tilde{s}_S$ at large $j_2 - j_1$ testify of the presence of long filaments.

\begin{figure*}
    \centering
    \resizebox{0.87\hsize}{!}{\includegraphics{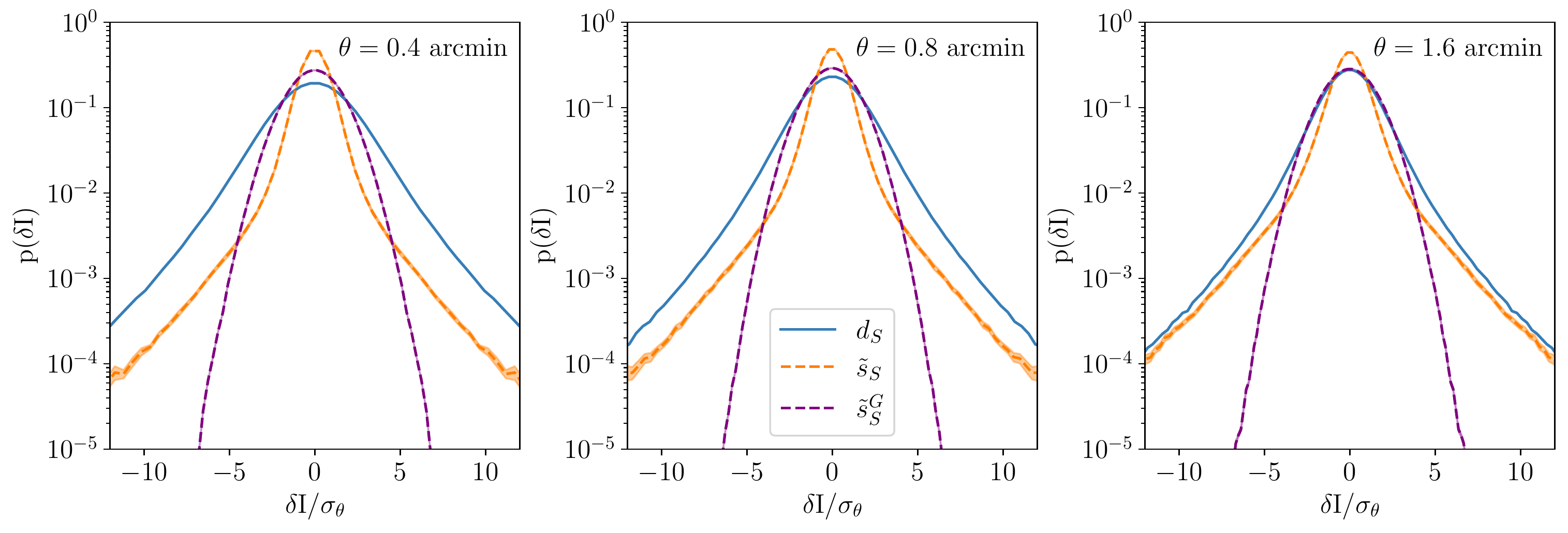}}
    \caption{Increment PDFs of the input and output dust maps for the component separation applied to \textit{Herschel} SPIRE maps at 250\micron. The increment $\rm{\delta I}$ of $d_S$, $\tilde{s}_S$, and $\tilde{s}_S^G$ are compared at $\theta =$ 0.4, 0.8, and 1.6 arcmin lags. The PDFs are displayed as a function of $\rm{\delta I / \sigma_\theta}$, where $\sigma_\theta$ is the standard deviation of the increments  of $\tilde{s}_S$ at lag $\theta$. The colored bands represent $\pm 1\sigma$ error-bars.}
    \label{fig:increment_PDF_Herschel_separation}
\end{figure*} 

\begin{figure*}
    \centering
    \resizebox{0.87\hsize}{!}{\includegraphics{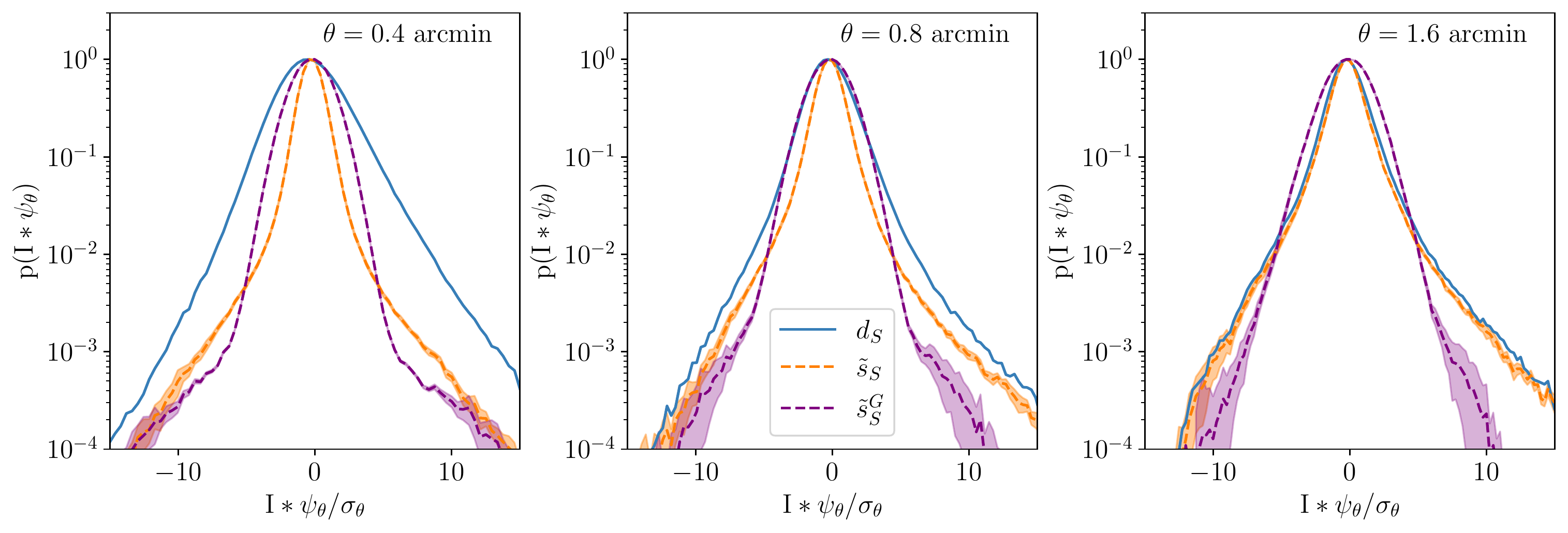}}
    \caption{Wavelet coefficents of the input and output dust maps for the component separation applied on the \textit{Herschel} Spider map. The PDFs of the real part of the wavelet coefficients of $d_S$, $\tilde{s}_S$, and $\tilde{s}_S^G$ are compared at scales $\theta =$ 0.4, 0.8, and $1.6^\prime$. They are displayed as a function of $\rm{I \ast \psi_\theta / \sigma_\theta}$ where $\psi_\theta$ is the wavelet at scale $\theta$ and $\sigma_\theta$ is the standard deviation of the wavelet coefficients. We employed the same oriented wavelets as those used to calculate WPH statistics. The PDFs are computed over all the pixels of the eight convolution maps obtained using the wavelets of different orientations. The colored bands represent $\pm 1\sigma$ error-bars.}
    \label{fig:wavelet_convolution_PDF_Herschel_separation}
\end{figure*} 

\subsection{Signatures of turbulence intermittency}
\label{subsec:intermittency}

Statistics of velocity increments derived from CO observations results have been theoretically interpreted as a signature of intermittency of interstellar turbulence \citep{Falgarone15}. The same result is theoretically expected for the dust polarization and the gas column density \citep{Momferratos15}.

Figure~\ref{fig:increment_PDF_Herschel_separation} presents the PDFs of the increment images of $\Tilde{s}_S$ shown in Fig.~\ref{fig:increment_maps_Herschel_separation}. The PDFs are roughly scale-invariant. They display non-Gaussian wings with no clear trend with the lag. This result contrasts with what has been reported for other observables. Indeed, the PDFs of increments computed for the gas velocity from spectroscopic CO observations \citep{hily-blant_dissipative_2008} and for the dust polarization from \textit{Planck} data \citep{regaldo-saint_blancard_new_2021}  show non-Gaussian wings that become increasingly prominent for decreasing lag. 

\subsection{Structure formation in the diffuse ISM}
\label{structure formation}

The formation of structure in the cold interstellar medium is thought to be driven by the interplay between the turbulent gas dynamics and thermal instability \citep{Kritsuk02,saury_structure_2014}. The statistics of dust maps provide observational insight. \citet{Mamd07} presented PDFs of wavelet convolutions of IRAS dust images. They have on all scales a pronounced skewness, which they relate to the non-linearity of the dynamical processes driving the formation of structures. The \textit{Herschel} data and the dust and CIB separation allow us to extend this analysis to smaller angular scales.   

Figure~\ref{fig:wavelet_convolution_PDF_Herschel_separation} shows PDFs of wavelet convolutions on angular scales from 0.4\arcmin\ to 1.6\arcmin, well below the IRAS $5^\prime$ resolution. The wavelets are those used for the WPH data analysis. The PDFs of $\Tilde{s}_S$ exhibits non-Gaussian wings, but smaller skewness than that of $d_S$. Table~\ref{tab:skewkurt} contains the skewness and kurtosis of wavelet convolutions at scales from 0.4\arcmin\ to 12.8\arcmin. These values describe the shape of the PDFs. Our values are not directly comparable to those computed by \citep{Mamd07} on IRAS maps because we use directional wavelets instead of isotropic wavelets.

\begin{table}[h!]
    \centering
    \begin{tabular}{|c|c|c|}
    \hline
    Scale & Skewness & Kurtosis \\
    \hline
    0.4\arcmin & 0.84 $\pm$ 0.08 & 29.97 $\pm$ 1.66 \\
    \hline
    0.8\arcmin & 1.43 $\pm$ 0.08 & 28.24 $\pm$ 1.79 \\
    \hline
    1.6\arcmin & 1.25 $\pm$ 0.08 & 18.21 $\pm$ 2.18 \\
    \hline
    3.2\arcmin & 1.20 $\pm$ 0.11 & 13.16 $\pm$ 2.06 \\
    \hline
    6.4\arcmin & 0.95 $\pm$ 0.09 & 7.34 $\pm$ 0.57 \\
    \hline
    12.8\arcmin & 1.22 $\pm$ 0.08 & 7.39 $\pm$ 0.43 \\
    \hline
    \end{tabular}
    \caption{\centering Skewness and kurtosis of the wavelet convolutions of the dust map at 0.4\arcmin, 0.8\arcmin, 1.6\arcmin, 3.2\arcmin, 6.4\arcmin\ , and 12.8\arcmin.}
    \label{tab:skewkurt}
\end{table}

\vspace{1cm}

\section{Conclusion}
\label{sec:summary}

We have made use of the distinct textures of the CIB and the dust emission on the sky to develop and apply a component separation method on \textit{Herschel} observations at a single frequency. The main results of our work are as follows.

The component separation problem on \textit{Herschel} data involve three components: the dust, the CIB, and the data noise. We reduce this problem to a single inverse problem in terms of WPH statistics and present an algorithm to solve it. The results of this algorithm are a WPH generative model of the dust emission and an output map which is a realization of the model correlated with the input map.

We build a set of realistic mock data using an \HI map as a template of CIB-free dust emission map. These data are used to validate the method. We show that we are able to retrieve the WPH statistics of the dust emission as well as non-Gaussian statistics used in astrophysics to characterize interstellar imaging data. Unlike methods that minimize the mean squared error in pixel space, we reproduce the power spectrum of the input map down to the smallest angular scales.

The method is applied to a \textit{Herschel} SPIRE observation of diffuse interstellar matter at 250\micron. We succeed in performing a statistical separation from observational data only at a single frequency by using non-Gaussian statistics. The power spectrum of the output map is well fitted by a power-law up to $k = 2 \ \text{arcmin}^{-1}$, where the dust signal represents $2 \%$ of the total power. The obtained slope is $-2.95 \pm 0.04$. 

We analyzed the non-Gaussian properties of the Spider dust emission on scales where the CIB signal is dominant. The component separation step is essential for characterizing the non-Gaussianity of the dust emission. Going beyond a standard power spectra analysis, we show that the non-Gaussian properties of the dust emission are not scale-invariant. The separated dust map reveals coherent structures at the smallest scales.
This work offers several perspectives for future work:
\begin{itemize}
    \item We underline that we use in this paper only one of the three wavelengths of \textit{Herschel} SPIRE. Thanks to the recent development of cross-WPH statistics \citep{regaldo-saint_blancard_generative_2022}, our method could be extended to a multi-channel component separation of the dust and CIB on \textit{Herschel} data with an aim to statistically characterize the dust SED.
    
    \item We could use external templates as an \HI\ observation for the dust. 
    
    \item It would also be useful to compare the gas and dust to test whether these two interstellar components are coupled on small angular scales for the cold and warm phases. This work could also be extended to other fields mapped with SPIRE. 
    
    \item Our statistical component separation could also be used on \textit{Planck} data to separate the dust and CIB up to larger angular scales on the sphere.
\end{itemize}  

\begin{acknowledgements}
      We thank F.~Levrier and J.-M.~Delouis, as well as S.~Mallat and R.~Morel for helpful discussions. We also thanks the anonymous referee for valuable comments that helped to improve the paper. Software: PyWPH \citep{regaldo-saint_blancard_new_2021}, PyWST \citep{regaldo-saint_blancard_statistical_2020}.
\end{acknowledgements}

\bibliographystyle{aa}

\begin{thebibliography}{65}
\expandafter\ifx\csname natexlab\endcsname\relax\def\natexlab#1{#1}\fi

\bibitem[{{Abergel} {et~al.}(1996){Abergel}, {Boulanger}, {Delouis}, {Dudziak},
  \& {Steindling}}]{Abergel96}
{Abergel}, A., {Boulanger}, F., {Delouis}, J.~M., {Dudziak}, G., \&
  {Steindling}, S. 1996, \aap, 309, 245

\bibitem[{Allys {et~al.}(2019)Allys, Levrier, Zhang, Colling,
  Regaldo-Saint~Blancard, Boulanger, Hennebelle, \& Mallat}]{allys_rwst_2019}
Allys, E., Levrier, F., Zhang, S., {et~al.} 2019, A\&A, 629, A115

\bibitem[{Allys {et~al.}(2020)Allys, Marchand, Cardoso, Villaescusa-Navarro,
  Ho, \& Mallat}]{allys_new_2020}
Allys, E., Marchand, T., Cardoso, J.~F., {et~al.} 2020, Phys. Rev. D, 102,
  103506

\bibitem[{{Andr{\'e}} {et~al.}(2010){Andr{\'e}}, {Men'shchikov}, {Bontemps},
  {K{\"o}nyves}, {Motte}, {Schneider}, {Didelon}, {Minier}, {Saraceno},
  {Ward-Thompson}, {di Francesco}, {White}, {Molinari}, {Testi}, {Abergel},
  {Griffin}, {Henning}, {Royer}, {Mer{\'\i}n}, {Vavrek}, {Attard},
  {Arzoumanian}, {Wilson}, {Ade}, {Aussel}, {Baluteau}, {Benedettini},
  {Bernard}, {Blommaert}, {Cambr{\'e}sy}, {Cox}, {di Giorgio}, {Hargrave},
  {Hennemann}, {Huang}, {Kirk}, {Krause}, {Launhardt}, {Leeks}, {Le Pennec},
  {Li}, {Martin}, {Maury}, {Olofsson}, {Omont}, {Peretto}, {Pezzuto}, {Prusti},
  {Roussel}, {Russeil}, {Sauvage}, {Sibthorpe}, {Sicilia-Aguilar}, {Spinoglio},
  {Waelkens}, {Woodcraft}, \& {Zavagno}}]{Andre10}
{Andr{\'e}}, P., {Men'shchikov}, A., {Bontemps}, S., {et~al.} 2010, \aap, 518,
  L102

\bibitem[{{B{\'e}thermin} {et~al.}(2013){B{\'e}thermin}, {Wang}, {Dor{\'e}},
  {Lagache}, {Sargent}, {Daddi}, {Cousin}, \& {Aussel}}]{Bethermin13}
{B{\'e}thermin}, M., {Wang}, L., {Dor{\'e}}, O., {et~al.} 2013, \aap, 557, A66

\bibitem[{Blagrave {et~al.}(2017)Blagrave, Martin, Joncas, Kothes, Stil,
  Miville-Deschênes, Lockman, \& Taylor}]{blagrave_dhigls_2017}
Blagrave, K., Martin, P.~G., Joncas, G., {et~al.} 2017, ApJ, 834, 126

\bibitem[{Boulanger {et~al.}(1996)Boulanger, Abergel, Bernard, Burton, Desert,
  Hartmann, Lagache, \& Puget}]{boulanger_dustgas_1996}
Boulanger, F., Abergel, A., Bernard, J.~P., {et~al.} 1996, A\&A, 312, 256

\bibitem[{Bruna \& Mallat(2013)}]{bruna2013invariant}
Bruna, J. \& Mallat, S. 2013, IEEE transactions on pattern analysis and machine
  intelligence, 35, 1872

\bibitem[{Bruna \& Mallat(2019)}]{Bruna2019}
Bruna, J. \& Mallat, S. 2019, Mathematical Statistics and Learning, 1, 257

\bibitem[{Burkhart(2021)}]{burkhart_diagnosing_2021}
Burkhart, B. 2021, PASP, 133, 102001

\bibitem[{{Cheng} \& {M{\'e}nard}(2021)}]{Cheng21}
{Cheng}, S. \& {M{\'e}nard}, B. 2021, arXiv e-prints, arXiv:2112.01288

\bibitem[{Chiang \& Ménard(2019)}]{chiang_extragalactic_2019}
Chiang, Y.-K. \& Ménard, B. 2019, The Astrophysical Journal, 870, 120

\bibitem[{Delouis {et~al.}(2022)Delouis, Allys, Gauvrit, \&
  Boulanger}]{delouis_non-gaussian_2022}
Delouis, J.~M., Allys, E., Gauvrit, E., \& Boulanger, F. 2022, A\&A, 668, A122

\bibitem[{{Falgarone} {et~al.}(2015){Falgarone}, {Momferratos}, \&
  {Lesaffre}}]{Falgarone15}
{Falgarone}, E., {Momferratos}, G., \& {Lesaffre}, P. 2015, in Astrophysics and
  Space Science Library, Vol. 407, Magnetic Fields in Diffuse Media, ed.
  A.~{Lazarian}, E.~M. {de Gouveia Dal Pino}, \& C.~{Melioli}, 227

\bibitem[{{Gautier} {et~al.}(1992){Gautier}, {Boulanger}, {Perault}, \&
  {Puget}}]{Gautier92}
{Gautier}, T.~N., I., {Boulanger}, F., {Perault}, M., \& {Puget}, J.~L. 1992,
  \aj, 103, 1313

\bibitem[{{Griffin} {et~al.}(2010){Griffin}, {Abergel}, {Abreu}, {Ade},
  {Andr{\'e}}, {Augueres}, {Babbedge}, {Bae}, {Baillie}, {Baluteau}, {Barlow},
  {Bendo}, {Benielli}, {Bock}, {Bonhomme}, {Brisbin}, {Brockley-Blatt},
  {Caldwell}, {Cara}, {Castro-Rodriguez}, {Cerulli}, {Chanial}, {Chen},
  {Clark}, {Clements}, {Clerc}, {Coker}, {Communal}, {Conversi}, {Cox},
  {Crumb}, {Cunningham}, {Daly}, {Davis}, {de Antoni}, {Delderfield}, {Devin},
  {di Giorgio}, {Didschuns}, {Dohlen}, {Donati}, {Dowell}, {Dowell}, {Duband},
  {Dumaye}, {Emery}, {Ferlet}, {Ferrand}, {Fontignie}, {Fox}, {Franceschini},
  {Frerking}, {Fulton}, {Garcia}, {Gastaud}, {Gear}, {Glenn}, {Goizel},
  {Griffin}, {Grundy}, {Guest}, {Guillemet}, {Hargrave}, {Harwit}, {Hastings},
  {Hatziminaoglou}, {Herman}, {Hinde}, {Hristov}, {Huang}, {Imhof}, {Isaak},
  {Israelsson}, {Ivison}, {Jennings}, {Kiernan}, {King}, {Lange}, {Latter},
  {Laurent}, {Laurent}, {Leeks}, {Lellouch}, {Levenson}, {Li}, {Li},
  {Lilienthal}, {Lim}, {Liu}, {Lu}, {Madden}, {Mainetti}, {Marliani}, {McKay},
  {Mercier}, {Molinari}, {Morris}, {Moseley}, {Mulder}, {Mur}, {Naylor},
  {Nguyen}, {O'Halloran}, {Oliver}, {Olofsson}, {Olofsson}, {Orfei}, {Page},
  {Pain}, {Panuzzo}, {Papageorgiou}, {Parks}, {Parr-Burman}, {Pearce},
  {Pearson}, {P{\'e}rez-Fournon}, {Pinsard}, {Pisano}, {Podosek}, {Pohlen},
  {Polehampton}, {Pouliquen}, {Rigopoulou}, {Rizzo}, {Roseboom}, {Roussel},
  {Rowan-Robinson}, {Rownd}, {Saraceno}, {Sauvage}, {Savage}, {Savini},
  {Sawyer}, {Scharmberg}, {Schmitt}, {Schneider}, {Schulz}, {Schwartz},
  {Shafer}, {Shupe}, {Sibthorpe}, {Sidher}, {Smith}, {Smith}, {Smith},
  {Spencer}, {Stobie}, {Sudiwala}, {Sukhatme}, {Surace}, {Stevens}, {Swinyard},
  {Trichas}, {Tourette}, {Triou}, {Tseng}, {Tucker}, {Turner}, {Vaccari},
  {Valtchanov}, {Vigroux}, {Virique}, {Voellmer}, {Walker}, {Ward}, {Waskett},
  {Weilert}, {Wesson}, {White}, {Whitehouse}, {Wilson}, {Winter}, {Woodcraft},
  {Wright}, {Xu}, {Zavagno}, {Zemcov}, {Zhang}, \& {Zonca}}]{Griffin10}
{Griffin}, M.~J., {Abergel}, A., {Abreu}, A., {et~al.} 2010, \aap, 518, L3

\bibitem[{{Hauser} \& {Dwek}(2001)}]{Hauser01}
{Hauser}, M.~G. \& {Dwek}, E. 2001, \araa, 39, 249

\bibitem[{Heiles(1984)}]{heiles_hi_1984}
Heiles, C. 1984, ApJ Supplement Series, 55, 585

\bibitem[{Heiles(1989)}]{heiles_magnetic_1989}
Heiles, C. 1989, ApJ, 336, 808

\bibitem[{{Hennebelle} \& {Inutsuka}(2019)}]{Hennebelle19}
{Hennebelle}, P. \& {Inutsuka}, S.-i. 2019, Frontiers in Astronomy and Space
  Sciences, 6, 5

\bibitem[{Hily-Blant {et~al.}(2008)Hily-Blant, Falgarone, \&
  Pety}]{hily-blant_dissipative_2008}
Hily-Blant, P., Falgarone, E., \& Pety, J. 2008, A\&A, 481, 367

\bibitem[{{Jewell}(2001)}]{Jewell01}
{Jewell}, J. 2001, \apj, 557, 700

\bibitem[{{Knox} {et~al.}(2001){Knox}, {Cooray}, {Eisenstein}, \&
  {Haiman}}]{Knox01}
{Knox}, L., {Cooray}, A., {Eisenstein}, D., \& {Haiman}, Z. 2001, \apj, 550, 7

\bibitem[{Knox(1995)}]{knox_perrins_1995}
Knox, R. 1995, Journal of Luminescence, 63, 163

\bibitem[{{Kritsuk} \& {Norman}(2002)}]{Kritsuk02}
{Kritsuk}, A.~G. \& {Norman}, M.~L. 2002, \apjl, 569, L127

\bibitem[{{Lagache} {et~al.}(2007){Lagache}, {Bavouzet}, {Fernandez-Conde},
  {Ponthieu}, {Rodet}, {Dole}, {Miville-Desch{\^e}nes}, \& {Puget}}]{Lagache07}
{Lagache}, G., {Bavouzet}, N., {Fernandez-Conde}, N., {et~al.} 2007, \apjl,
  665, L89

\bibitem[{{Lei} \& {Clark}(2022)}]{Lei_Clark22}
{Lei}, M. \& {Clark}, S.~E. 2022, arXiv e-prints, arXiv:2212.06182

\bibitem[{Lenz {et~al.}(2019)Lenz, Doré, \& Lagache}]{lenz_large-scale_2019}
Lenz, D., Doré, O., \& Lagache, G. 2019, ApJ, 883, 75

\bibitem[{{Lockman} {et~al.}(1986){Lockman}, {Jahoda}, \&
  {McCammon}}]{Lockman86}
{Lockman}, F.~J., {Jahoda}, K., \& {McCammon}, D. 1986, \apj, 302, 432

\bibitem[{Lombardi {et~al.}(2015)Lombardi, Alves, \&
  Lada}]{lombardi_molecular_2015}
Lombardi, M., Alves, J., \& Lada, C.~J. 2015, A\&A, 576, L1

\bibitem[{{Mak} {et~al.}(2017){Mak}, {Challinor}, {Efstathiou}, \&
  {Lagache}}]{Mak17}
{Mak}, D. S.~Y., {Challinor}, A., {Efstathiou}, G., \& {Lagache}, G. 2017,
  \mnras, 466, 286

\bibitem[{Mallat(2012)}]{mallat_group_2012}
Mallat, S. 2012, Communications on Pure and Applied Mathematics, 65, 1331

\bibitem[{{Maniyar} {et~al.}(2018){Maniyar}, {B{\'e}thermin}, \&
  {Lagache}}]{Maniyar18}
{Maniyar}, A.~S., {B{\'e}thermin}, M., \& {Lagache}, G. 2018, \aap, 614, A39

\bibitem[{{Marchal} \& {Martin}(2023)}]{Marchal23}
{Marchal}, A. \& {Martin}, P.~G. 2023, \apj, submitted.

\bibitem[{Marchal {et~al.}(2021)Marchal, Martin, \&
  Gong}]{marchal_resolving_2021}
Marchal, A., Martin, P.~G., \& Gong, M. 2021, ApJ, 921, 11

\bibitem[{Marchal \& Miville-Deschênes(2021)}]{marchal_thermal_2021}
Marchal, A. \& Miville-Deschênes, M.~A. 2021, ApJ, 908, 186

\bibitem[{Marchal {et~al.}(2019)Marchal, Miville-Deschênes, Orieux, Gac,
  Soussen, Lesot, d'Allonnes, \& Salomé}]{marchal_rohsa_2019}
Marchal, A., Miville-Deschênes, M.~A., Orieux, F., {et~al.} 2019, A\&A, 626,
  A101

\bibitem[{Meyer \& Roth(1991)}]{meyer_discovery_1991}
Meyer, D.~M. \& Roth, K.~C. 1991, ApJ, 376, L49

\bibitem[{{Miville-Desch{\^e}nes} {et~al.}(2007){Miville-Desch{\^e}nes},
  {Lagache}, {Boulanger}, \& {Puget}}]{Mamd07}
{Miville-Desch{\^e}nes}, M.~A., {Lagache}, G., {Boulanger}, F., \& {Puget},
  J.~L. 2007, \aap, 469, 595

\bibitem[{{Miville-Desch{\^e}nes} {et~al.}(2002){Miville-Desch{\^e}nes},
  {Lagache}, \& {Puget}}]{Mamd02}
{Miville-Desch{\^e}nes}, M.~A., {Lagache}, G., \& {Puget}, J.~L. 2002, \aap,
  393, 749

\bibitem[{{Miville-Desch{\^e}nes} {et~al.}(2010){Miville-Desch{\^e}nes},
  {Martin}, {Abergel}, {Bernard}, {Boulanger}, {Lagache}, {Anderson},
  {Andr{\'e}}, {Arab}, {Baluteau}, {Blagrave}, {Bontemps}, {Cohen},
  {Compiegne}, {Cox}, {Dartois}, {Davis}, {Emery}, {Fulton}, {Gry}, {Habart},
  {Huang}, {Joblin}, {Jones}, {Kirk}, {Lim}, {Madden}, {Makiwa}, {Menshchikov},
  {Molinari}, {Moseley}, {Motte}, {Naylor}, {Okumura}, {Pinheiro
  Gon{\c{c}}alves}, {Polehampton}, {Rod{\'o}n}, {Russeil}, {Saraceno},
  {Schneider}, {Sidher}, {Spencer}, {Swinyard}, {Ward-Thompson}, {White}, \&
  {Zavagno}}]{Mamd10}
{Miville-Desch{\^e}nes}, M.~A., {Martin}, P.~G., {Abergel}, A., {et~al.} 2010,
  \aap, 518, L104

\bibitem[{Miville-Deschênes {et~al.}(2016)Miville-Deschênes, Duc, Marleau,
  Cuillandre, Didelon, Gwyn, \& Karabal}]{miville-deschenes_probing_2016}
Miville-Deschênes, M.~A., Duc, P.~A., Marleau, F., {et~al.} 2016, A\&A, 593,
  A4

\bibitem[{{Momferratos}(2015)}]{Momferratos15}
{Momferratos}, G. 2015, PhD thesis, Universit\'e Paris-Sud

\bibitem[{{Oliver} {et~al.}(2012){Oliver}, {Bock}, {Altieri}, {Amblard},
  {Arumugam}, {Aussel}, {Babbedge}, {Beelen}, {B{\'e}thermin}, {Blain},
  {Boselli}, {Bridge}, {Brisbin}, {Buat}, {Burgarella},
  {Castro-Rodr{\'\i}guez}, {Cava}, {Chanial}, {Cirasuolo}, {Clements},
  {Conley}, {Conversi}, {Cooray}, {Dowell}, {Dubois}, {Dwek}, {Dye}, {Eales},
  {Elbaz}, {Farrah}, {Feltre}, {Ferrero}, {Fiolet}, {Fox}, {Franceschini},
  {Gear}, {Giovannoli}, {Glenn}, {Gong}, {Gonz{\'a}lez Solares}, {Griffin},
  {Halpern}, {Harwit}, {Hatziminaoglou}, {Heinis}, {Hurley}, {Hwang}, {Hyde},
  {Ibar}, {Ilbert}, {Isaak}, {Ivison}, {Lagache}, {Le Floc'h}, {Levenson},
  {Faro}, {Lu}, {Madden}, {Maffei}, {Magdis}, {Mainetti}, {Marchetti},
  {Marsden}, {Marshall}, {Mortier}, {Nguyen}, {O'Halloran}, {Omont}, {Page},
  {Panuzzo}, {Papageorgiou}, {Patel}, {Pearson}, {P{\'e}rez-Fournon}, {Pohlen},
  {Rawlings}, {Raymond}, {Rigopoulou}, {Riguccini}, {Rizzo}, {Rodighiero},
  {Roseboom}, {Rowan-Robinson}, {S{\'a}nchez Portal}, {Schulz}, {Scott},
  {Seymour}, {Shupe}, {Smith}, {Stevens}, {Symeonidis}, {Trichas}, {Tugwell},
  {Vaccari}, {Valtchanov}, {Vieira}, {Viero}, {Vigroux}, {Wang}, {Ward},
  {Wardlow}, {Wright}, {Xu}, \& {Zemcov}}]{Oliver12}
{Oliver}, S.~J., {Bock}, J., {Altieri}, B., {et~al.} 2012, \mnras, 424, 1614

\bibitem[{{Planck 2013 results. XVIII}(2014)}]{Planck_XVIII_2013}
{Planck 2013 results. XVIII}. 2014, \aap, 571, A18

\bibitem[{{Planck Collaboration}(2014)}]{PlanckXXX_CIB}
{Planck Collaboration}. 2014, \aap, 571, A30

\bibitem[{{Planck Collaboration}(2016{\natexlab{a}})}]{PIPXLVIII}
{Planck Collaboration}. 2016{\natexlab{a}}, \aap, 596, A109

\bibitem[{{Planck Collaboration}(2016{\natexlab{b}})}]{Planck_XXX}
{Planck Collaboration}. 2016{\natexlab{b}}, \aap, 586, A133

\bibitem[{{Planck Collaboration}(2020)}]{Planck2018_IV}
{Planck Collaboration}. 2020, \aap, 641, A4

\bibitem[{Remazeilles {et~al.}(2011)Remazeilles, Delabrouille, \&
  Cardoso}]{remazeilles_foreground_2011}
Remazeilles, M., Delabrouille, J., \& Cardoso, J.~F. 2011, MNRAS, 418, 467

\bibitem[{{Robitaille} {et~al.}(2019){Robitaille}, {Motte}, {Schneider},
  {Elia}, \& {Bontemps}}]{Robitaille19}
{Robitaille}, J.~F., {Motte}, F., {Schneider}, N., {Elia}, D., \& {Bontemps},
  S. 2019, \aap, 628, A33

\bibitem[{Régaldo-Saint~Blancard {et~al.}(2022)Régaldo-Saint~Blancard, Allys,
  Auclair, Boulanger, Eickenberg, Levrier, Vacher, \&
  Zhang}]{regaldo-saint_blancard_generative_2022}
Régaldo-Saint~Blancard, B., Allys, E., Auclair, C., {et~al.} 2022, ApJ, 943, 9

\bibitem[{Régaldo-Saint~Blancard {et~al.}(2021)Régaldo-Saint~Blancard, Allys,
  Boulanger, Levrier, \& Jeffrey}]{regaldo-saint_blancard_new_2021}
Régaldo-Saint~Blancard, B., Allys, E., Boulanger, F., Levrier, F., \& Jeffrey,
  N. 2021, A\&A, 649, L18

\bibitem[{Régaldo-Saint~Blancard {et~al.}(2020)Régaldo-Saint~Blancard,
  Levrier, Allys, Bellomi, \&
  Boulanger}]{regaldo-saint_blancard_statistical_2020}
Régaldo-Saint~Blancard, B., Levrier, F., Allys, E., Bellomi, E., \& Boulanger,
  F. 2020, A\&A, 642, A217

\bibitem[{Saury {et~al.}(2014)Saury, Miville-Deschênes, Hennebelle, Audit, \&
  Schmidt}]{saury_structure_2014}
Saury, E., Miville-Deschênes, M.~A., Hennebelle, P., Audit, E., \& Schmidt, W.
  2014, A\&A, 567, A16

\bibitem[{{Saydjari} {et~al.}(2021){Saydjari}, {Portillo}, {Slepian},
  {Kahraman}, {Burkhart}, \& {Finkbeiner}}]{Saydjari21}
{Saydjari}, A.~K., {Portillo}, S. K.~N., {Slepian}, Z., {et~al.} 2021, \apj,
  910, 122

\bibitem[{Scott {et~al.}(1994)Scott, Srednicki, \& White}]{scott_sample_1994}
Scott, D., Srednicki, M., \& White, M. 1994, ApJ, 421, L5

\bibitem[{{Serra} {et~al.}(2014){Serra}, {Lagache}, {Dor{\'e}}, {Pullen}, \&
  {White}}]{Serra14}
{Serra}, P., {Lagache}, G., {Dor{\'e}}, O., {Pullen}, A., \& {White}, M. 2014,
  \aap, 570, A98

\bibitem[{{Shirley} {et~al.}(2021){Shirley}, {Duncan}, {Campos Varillas},
  {Hurley}, {Ma{\l}ek}, {Roehlly}, {Smith}, {Aussel}, {Bakx}, {Buat},
  {Burgarella}, {Christopher}, {Duivenvoorden}, {Eales}, {Efstathiou},
  {Gonz{\'a}lez Solares}, {Griffin}, {Jarvis}, {Faro}, {Marchetti}, {McCheyne},
  {Papadopoulos}, {Penner}, {Pons}, {Prescott}, {Rigby}, {Rottgering},
  {Saxena}, {Scudder}, {Vaccari}, {Wang}, \& {Oliver}}]{Shirley21}
{Shirley}, R., {Duncan}, K., {Campos Varillas}, M.~C., {et~al.} 2021, \mnras,
  507, 129

\bibitem[{Taank {et~al.}(2022)Taank, Marchal, Martin, \&
  Vujeva}]{taank_mapping_2022}
Taank, M., Marchal, A., Martin, P.~G., \& Vujeva, L. 2022, ApJ, 937, 81

\bibitem[{{V{\'a}zquez-Semadeni} {et~al.}(2000){V{\'a}zquez-Semadeni}, {Gazol},
  \& {Scalo}}]{Vazquez00}
{V{\'a}zquez-Semadeni}, E., {Gazol}, A., \& {Scalo}, J. 2000, \apj, 540, 271

\bibitem[{{Viero} {et~al.}(2013){Viero}, {Wang}, {Zemcov}, {Addison},
  {Amblard}, {Arumugam}, {Aussel}, {B{\'e}thermin}, {Bock}, {Boselli}, {Buat},
  {Burgarella}, {Casey}, {Clements}, {Conley}, {Conversi}, {Cooray}, {De
  Zotti}, {Dowell}, {Farrah}, {Franceschini}, {Glenn}, {Griffin},
  {Hatziminaoglou}, {Heinis}, {Ibar}, {Ivison}, {Lagache}, {Levenson},
  {Marchetti}, {Marsden}, {Nguyen}, {O'Halloran}, {Oliver}, {Omont}, {Page},
  {Papageorgiou}, {Pearson}, {P{\'e}rez-Fournon}, {Pohlen}, {Rigopoulou},
  {Roseboom}, {Rowan-Robinson}, {Schulz}, {Scott}, {Seymour}, {Shupe}, {Smith},
  {Symeonidis}, {Vaccari}, {Valtchanov}, {Vieira}, {Wardlow}, \&
  {Xu}}]{Viero13}
{Viero}, M.~P., {Wang}, L., {Zemcov}, M., {et~al.} 2013, \apj, 772, 77

\bibitem[{Wiener(1949)}]{wiener_extrapolation_1949}
Wiener, N. 1949, Cambridge, MA: MIT Press

\bibitem[{{Yahia} {et~al.}(2021){Yahia}, {Schneider}, {Bontemps}, {Bonne},
  {Attuel}, {Dib}, {Ossenkopf-Okada}, {Turiel}, {Zebadua}, {Elia}, {Kumar
  Maji}, {Schmitt}, \& {Robitaille}}]{Yahia21}
{Yahia}, H., {Schneider}, N., {Bontemps}, S., {et~al.} 2021, \aap, 649, A33

\bibitem[{Zhang \& Mallat(2019)}]{zhang_maximum_2019}
Zhang, S. \& Mallat, S. 2019, Maximum {Entropy} {Models} from {Phase}
  {Harmonic} {Covariances}, Tech. rep.

\end{thebibliography}

\appendix

\section{Mathematical notations}
\label{app:notations}

\begin{table}[h!]
\centering

\begin{tabular}{| m{1.1cm} | m{7.2cm} |}
  \hline
  Notation & \hspace{2cm} Description \\
  \hline 
   \multicolumn{2}{|c|}{\vspace{-0.32cm}} \\
  \multicolumn{2}{|c|}{\vspace{0.01cm}{{\emph Herschel} SPIRE observations (Sect.~\ref{sec:obs})}}  \\
  \hline
  $d_L$ & Map of the Lockman Hole (LH) field \\
  $d_{\mathrm{PSF}}$ & Map of the \textit{Herschel} point spread function  \\
  $d_N$ & Noise map of the Spider field \\
  $d_{S1},d_{S2}$ & Two independent maps of the Spider field \\ 
  $d_S$ & Mean of $d_{S1}$ and $d_{S2}$  \\
  \hline
  \multicolumn{2}{|c|}{\vspace{-0.32cm}} \\
  \multicolumn{2}{|c|}{\vspace{0.01cm}{Validation of the dust and CIB sep. on mock data (Sect.~\ref{sec:validation})}} \\
  \hline
 $c^{\prime \ m,i}_S$ & Set of contamination maps: CIB + Spider data noise \\
  $\tilde{c}^{\prime \ m,i}_S$ & Set of output estimates of $c^{\prime \ m,i}_S$ from comp. sep.\\
  $d^{m,i}_S$ & Set of data maps obtained adding $c^{\prime \ m,i}_S$ to $s^m_S$  \\
  $s^m_S$ & Input dust map built from \HI\ observation \\
  $\tilde{s}^{m,i}_S$ & Set of output estimates of $s^m_S$ from comp. sep. \\
  \hline
  \multicolumn{2}{|c|}{\vspace{-0.32cm}} \\
  \multicolumn{2}{|c|}{\vspace{0.01cm}{Comp. sep. on {\emph Herschel} SPIRE \emph{Spider} field (Sects.~\ref{sec:application} and \ref{sec:non-Gaussian analysis})}} \\
  \hline
  $c^{\prime}_L$ & Contamination map: LH map + Spider data noise  \\
  $c^{\prime}_S$ & Spider contamination map \\
  $\tilde{c}^{\prime}_S$ & Output estimate of $c^{\prime}_S$ from component separation \\
  $\tilde{s}_S$ & Output dust map from component separation  \\
  $\tilde{s}_S^G$ & Gaussian realization of $\tilde{s}_S$ after phase rand. \\
  \hline
\end{tabular}

\caption{\centering Summary of the main mathematical notations.}
\label{tab:notations}
\end{table}

\section{WPH statistics}
\label{app:WPH_statistics}

We present our construction of the WPH statistics and explain how it can be used as a generative model. This brief overview is directly inspired by \citet{regaldo-saint_blancard_new_2021}.

\subsection{Definition}
\label{app:WPH_definition}

The construction of the WPH statistics follows two main steps: i) a multi-scale decomposition of the process under study into its different components and ii) the characterization of the interaction between its different scales.

\begin{figure}[h!]
    \centering
    \includegraphics[width=\hsize]{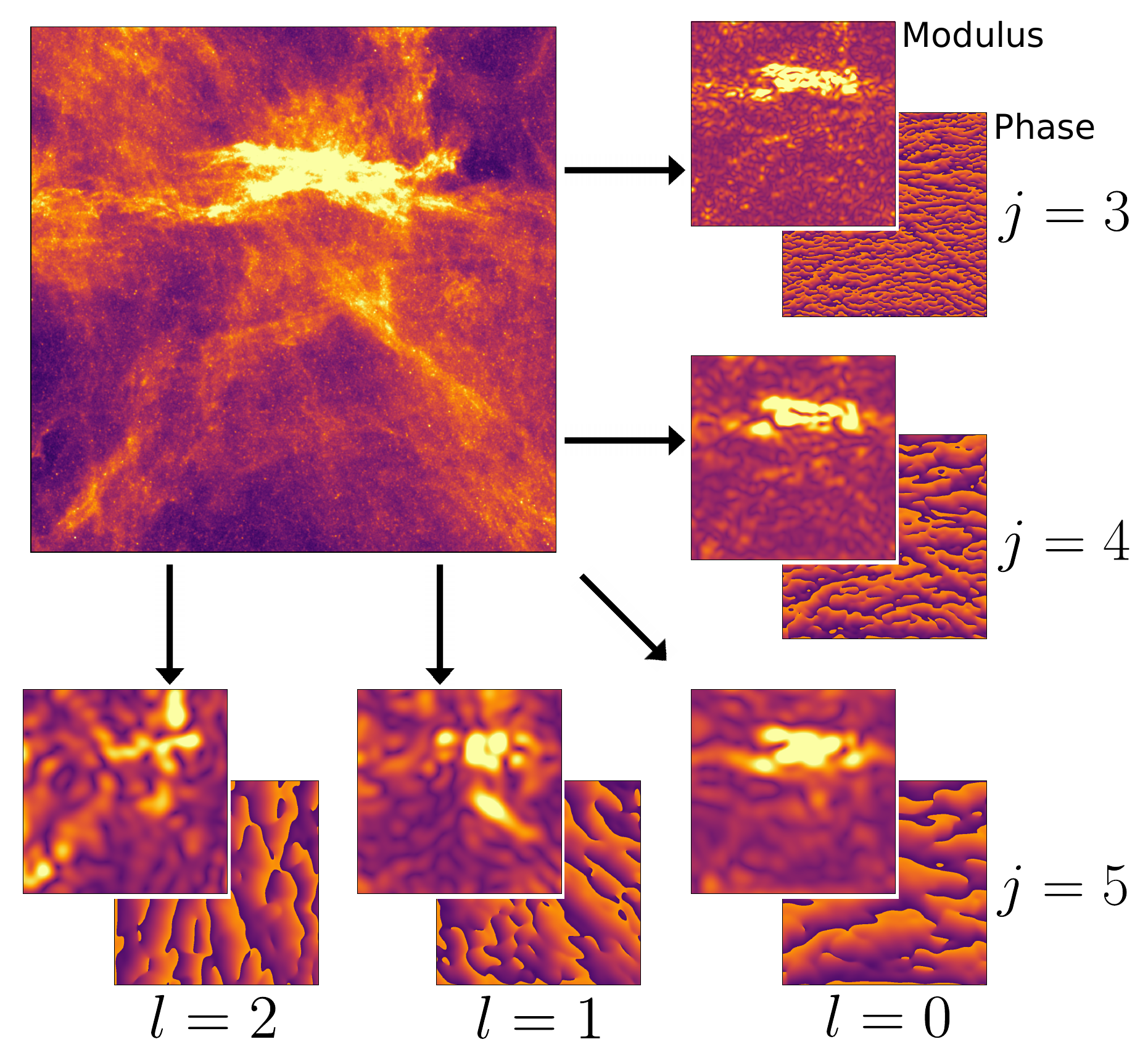}
    \caption{Modulus and phase of the convolutions of $d_S$ with $\psi_{3,0}$, $\psi_{4,0}$, $\psi_{5,0}$, $\psi_{5,1}$, and $\psi_{5,2}$. In this example, we set $L=4$.}
    \label{fig:WPH_decomposition}
\end{figure}

The first step consists of performing a multi-scale decomposition using a wavelet transform. The wavelets we use are bump steerable wavelets \citep{zhang_maximum_2019}. These wavelets $\psi_{j,l}(\vec{x})$ are indexed by two integers, $j \in [0,J-1]$ and $l \in [0,L-1]$, which define the oriented scale they characterize: $\psi_{j,l}(\vec{x})$ is the wavelet used to probe the $2^{j}$ characteristic scale at an angle $l\frac{2\pi}{L}$ from the reference axis. The wavelet transform of a process $\rho(\vec{x})$ is defined as the set of its convolutions $\rho*\psi_{j,l}(\vec{x})$ with wavelets at all oriented scales. It performs a multi-scale decomposition of $\rho$, decomposing its structures on the different scales probed by the wavelets: indeed, by convolving an image with a wavelet, we make a local filtering on the wavelet band-pass. This decomposition is illustrated in Fig.~\ref{fig:WPH_decomposition}, which shows the modulus and phase of convolutions of the $d_S$ map with several wavelets. For a given wavelet $\psi_{j,l}$, the modulus of the convolution $d_S * \psi_{j,l}$ highlights the structures of $d_S$ for a scale of $2^{j}$ and orientation of $l\frac{2\pi}{L}$.

The second step of this transform is to characterise the interaction between the different scales of the process, $\rho,$ under study. A natural way to characterize such an interaction is to compute the covariances between the different $\rho*\psi_{j,l}(\vec{x})$ terms of the wavelet transform. 
However, such covariances are not able to characterize non-Gaussian features. Indeed, we can show (see, e.g., \citet{allys_new_2020}) that:
\begin{equation}
     \text{Cov}(\rho*\psi_1, \rho*\psi_2) 
     = \int S (\vec{k}) 
     \ \hat \psi_1(\vec{k})
     \ \overline{\hat \psi_2(\vec{k})}
     \ \mathrm{d} \vec{k},
     \label{eq:cov}
\end{equation}
where $S(\vec{k})$ is the power spectrum of $\rho$, which only characterize each wavenumber, $\vec{k,}$ independently. This shows that in order to characterize coupling between scales, it is necessary to introduce non-linearities.

The second step to construct WPH statistics thus consists in introducing non-linearities in order to characterize interactions between scales. Indeed, if we want to get an information from the covariance between two fields, they must have common frequencies. We can take the modulus of the different wavelet convolutions, but also use a non-linear operator to capture information about phase alignement between different scales. This operator is the Phase Harmonics operator, which is defined as: 
\begin{equation}
    \forall z \in \mathbb{C}, \ \forall p \in \mathbb{N}, \ \ \ [z]^{p} = |z| \ e^{i \ arg(z) \times p}.
\end{equation}
For $p \neq 0$, this operator multiplies the complex phase of a field by a constant integer, computing the harmonics of its phase. This operator makes the phase of the two filtered images vary at the same spatial frequency in order to characterize the statistical phase alignment by means of a covariance.

We can now construct different "WPH moments," by computing covariance of phase harmonics of wavelet convolutions which share common frequencies, namely, whose spectral supports overlap. The WPH moments depend on a translation vector, $\tau,$ that allows us to increase the spectral resolution. The general expression of WPH moments is:
\begin{equation}
    C_{j_1,l_1,p_1,j_2,l_2,p_2}(\vec{\tau})=Cov([\rho * \psi_{j_1,l_1}]^{p_1}(\vec{x}),[\rho * \psi_{j_2,l_2}]^{p_2}(\vec{x}+\vec{\tau})).
\end{equation}
The number of translation vectors considered is $1 + 8 \Delta_n$. We note that due to the choice of the harmonics $p_1$ and $p_2$, there are different ways to couple a pair of scale.

The summary statistics we will use are WPH statistics, which are built from a set of WPH moments. In this work, we take a set of moments defined in \citet{regaldo-saint_blancard_generative_2022}. This is a small modification of \citet{allys_new_2020}, in which the set of moments has been chosen to construct a generative model of the large-scale structure matter density field \citep{allys_new_2020}. With $J=7$, $L=4,$ and $\Delta_n = 2,$ it gives a set of 3441 moments. To describe these moments, we will follow the notations of this paper and define several types of moments: if the two scales are equal, we use the letter $S,$ and if not, we use a $C$; the type of moment is identified by their respective harmonics $p_1$ and $p_2$. 

The harmonics, $p,$ multiply the frequency of a signal by a factor, $p$. As we want similar frequencies in order to have meaningful moments, several choices of harmonics are possible for a given pair of scales. First, we can take $(p_1,p_2)=(0,0),$ which will lead to common frequencies even if the scales are different. We can also take $(p_1,p_2)=(0,1)$ which will lead to common frequencies if $j_2 > j_1$. Finally, if we want to keep the phases and to take different scales, we have to choose $p_1$ and $p_2$ such that $j_1 p_1 \sim j_2 p_2$, and the simplest choice is then to take $(p_1,p_2)=(1,\frac{j_1}{j_2})$. For $j_1=j_2$, it boils down to $(p_1,p_2)=(1,1)$ which leads to the power spectrum-like moments, which we refer to the $S^{11}$ moments (see Eq.~\ref{eq:cov}). For each type of moments, the particular moment is then labelled by the characteristic spatial frequencies probed (one for $S$ term, two for $C$ terms), indexed by their couple $(j,l)$. For example, $C^{01}_{j_1,l_1,j_2,l_2}$ is the moment whose expression is $\text{Cov} \large( |\rho * \psi_{j_1,l_1}| (\vec{x}), (\rho * \psi_{j_2,l_2}) (\vec{x}) \large)$. \\

We will use six types of moments: $S^{11}$,\ $S^{00}$,\ $S^{01}$,\ $C^{\text{phase}}$,\ $C^{00}$, and $C^{01}$, whose expressions are as follows:
\begin{eqnarray}
\begin{aligned}
&S^{11}_{j,l}(\vec{\tau})= \text{Cov} \large( (\rho * \psi_{j,l}) (\vec{x}), (\rho * \psi_{j,l}) (\vec{x}+\vec{\tau}) \large), \\
&S^{00}_{j,l}(\vec{\tau})= \text{Cov} \large( |\rho * \psi_{j,l}| (\vec{x}), |\rho * \psi_{j,l}| (\vec{x}+\vec{\tau}) \large), \\
&S^{01}_{j,l}(\vec{\tau})= \text{Cov} \large( |\rho * \psi_{j,l}| (\vec{x}), (\rho * \psi_{j,l}) (\vec{x}+\vec{\tau}) \large), \\
&C^{\text{phase}}_{j_1,l_1,j_2,l_2}(\vec{\tau})= \text{Cov} \large( (\rho * \psi_{j_1,l_1}) (\vec{x}), [\rho * \psi_{j_2,l_2}]^{j_1 / j_2}(\vec{x}+\vec{\tau}) \large), \\
&C^{00}_{j_1,l_1,j_2,l_2}(\vec{\tau})= \text{Cov} \large( |\rho * \psi_{j_1,l_1}| (\vec{x}), |\rho * \psi_{j_2,l_2}| (\vec{x}+\vec{\tau}) \large), \\
&C^{01}_{j_1,l_1,j_2,l_2}(\vec{\tau})= \text{Cov} \large( |\rho * \psi_{j_1,l_1}| (\vec{x}), (\rho * \psi_{j_2,l_2}) (\vec{x}+\vec{\tau}) \large).
\end{aligned}
\end{eqnarray}
All these moments depend on the amplitude of the image power spectrum, but we want to have coefficients describing the non-Gaussian features only. Using Eq.~\ref{eq:cov}, we can show that the power spectrum information is contained in the $S^{11}$ moments. Then, by normalizing all the moments by the $S^{11}$ and the $S^{00}$ ones, we get coefficients that describe non-Gaussianity independently of the power spectrum. For example, the expressions of the normalized $C^{01}$ moments are $C^{01}_{j_1,l_1,j_2,l_2} / \sqrt{S^{00}_{j_1,l_1} S^{11}_{j_2,l_2}}$.

\subsection{WPH generative model}
\label{app:generative}

Generative models can be constructed from the WPH statistics of a given process. Here, this was done within the framework of maximum entropy models, in a microcanonical approach, which boils down to constructing the most general probability distribution under the WPH constraints~\citep{Bruna2019}. For a given statistical process, $X$, the WPH statistics allow to obtain new realizations, $X_i$, from a distribution estimated on an observation, $X_0$. Starting from a white noise, we perform a gradient descent in pixel space in order to reproduce the WPH statistics of $X_0$. Hereafter, we call this process a synthesis.

Many physical processes having power spectrum that vary over several orders of magnitude, their WPH statistics have very different values from one scale to another. This is a problem for the gradient descent because it gives an important weight to some scales, whereas others could be very little constrained. To prevent this problem, we normalize the WPH operator such that each WPH coefficient, which characterizes the coupling between two scales, is divided by the square root of the product of the power spectra of $X_0$ at the corresponding scales. The reference map, $m$ (defined in Sect.~\ref{subsec:algorithm}) is then set in this case to $X_0$.

On a practical level, a synthesis is done using a gradient descent on a map $u$ (where $u_0$ is a white noise) to minimize the loss $\displaystyle L_{syn}(u) = ||\Phi(u) - \Phi(X_0)||^2$, where $\Phi$ is the WPH operator. The result of the optimization is a new realization of the unkown process $X$ estimated on $X_0$, fully independent of $X_0$. \citet{allys_new_2020} showed that the WPH generative model reproduces usual non-Gaussian statistics in cosmology up to 1-10\%. The ability to build a non-Gaussian generative model is necessary for our component separation algorithm. 

\section{\HI map}
\label{app:HI_map}
The angular resolution of the ST interferometric data used as a spatial template of Galactic emission uncontaminated by the CIB is  0.91\arcmin and the pixel size is 18\arcsec.
The DHIGLS DF product has the full range of spatial frequencies, obtained by
a rigorous combination of the ST interferometric and GBT single dish data
\citep[see Section~5 in][]{blagrave_dhigls_2017}.
At small scales, \IHI\ is highly affected by noise, resulting in an increase of power at high $k$ \citep[][see their Figure 22]{blagrave_dhigls_2017}. The statistical properties of this noise is complex and can potentially alter the quality of the separation performed as a validation test. 
Because \IHI\ is only used as a template to generate a CIB-free SPIRE mock observation, we performed a denoising of the \HI\ data cube prior integration along the velocity axis (from -9.7 to 26.5 km/s) to obtain a map of \IHI\ in the Spider region with reduced data noise.

This was accomplished using the {\tt ROHSA} algorithm \citep{marchal_rohsa_2019}. While earlier applications were dedicated to phase separation \citep[e.g., ][]{marchal_resolving_2021,marchal_thermal_2021,taank_mapping_2022}, here we make use of the spatial regularization of {\tt ROHSA} to obtain a spatially coherent model of the data, which concurrently reduces noise. 
We first experimented with the decomposition of the DF dataset presented in \citet{Marchal23} that provides a full encoding of the data with $N=6$ Gaussians and the hyper-parameters $\lambda_{\ab}=\lambda_{\mub}=\lambda_{\sigmab}=\lambda_{\sigmab'}=10$, which control the strength of the regularization. Upon inspecting the resulting model, we found that the spatially correlated noise still dominates the signal on small scales in the \IHI\ map.
To overcome this limitation, we performed a new decomposition using higher hyper-parameters $\lambda_{\ab}=\lambda_{\mub}=\lambda_{\sigmab}=\lambda_{\sigmab'}=1000$ to control the spatial regularization and intentionally reduce the noise even further. The resulting model does not encode the signal fully (due to the strong regularization), which translates into a chi-square map that is not as uniform as the original decomposition. We emphasize that the goal of this procedure is solely to generate a realistic mock observation of Galactic dust based on existing high-resolution \HI\ data of the Spider, but \IHI\ is not a component of our wavelet-phase harmonic-based method.
\section{WPH statistics of the separation results on mock and \textit{Herschel} data}
\label{app:WPH_validation}

\begin{figure}[h!]
    \centering
    \includegraphics[width=\hsize]{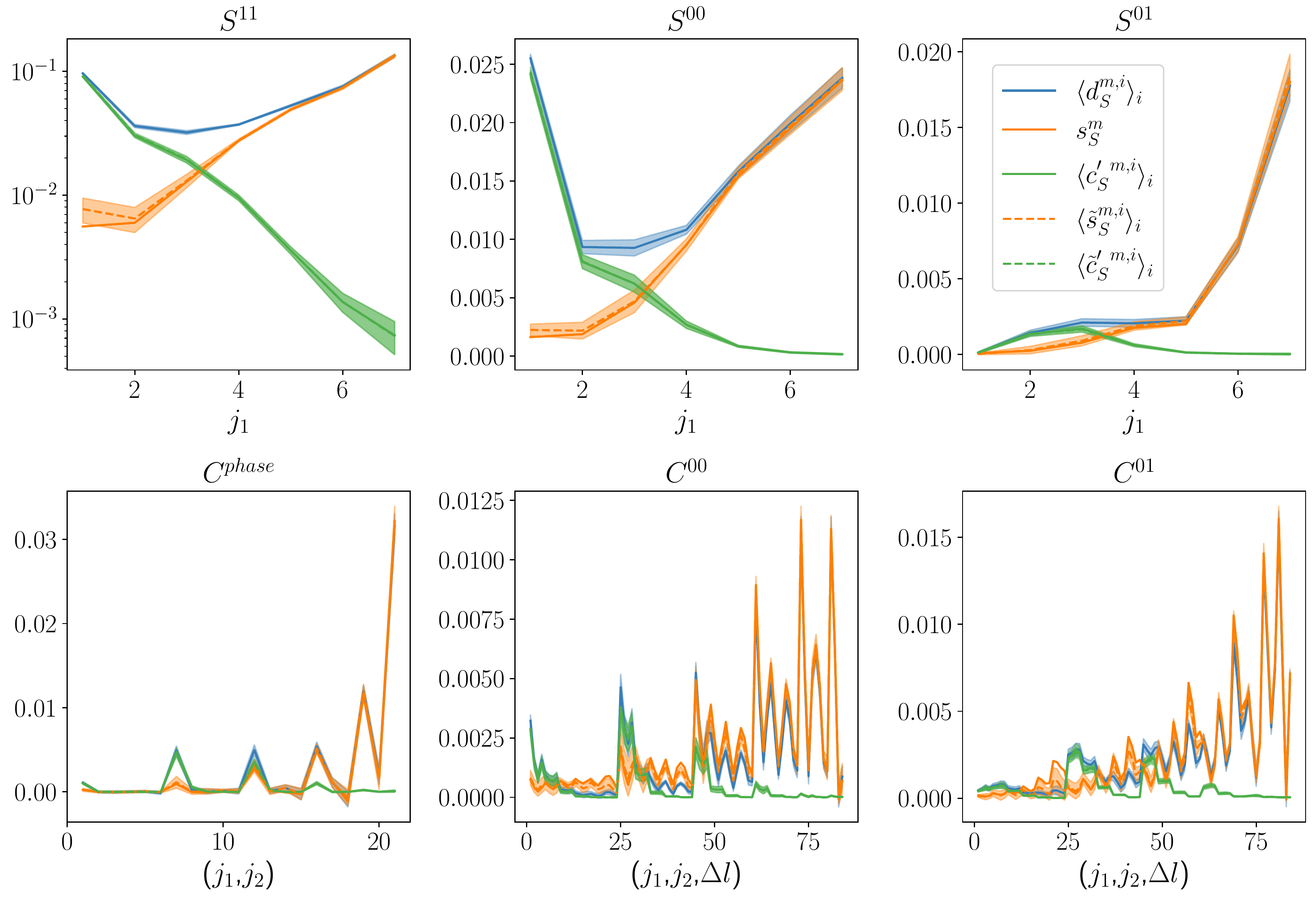}
    \caption{WPH statistics of the input and output maps for the component separation applied on the mock data. The WPH statistics of $d_S^{m,i}$, $s_S^m$, $c^{\prime \ m,i}_{S}$, $\tilde{s}_S^{m,i}$ and $\tilde{c}_S^{\prime \ m,i}$ are compared. The $\langle \cdot \rangle_i$ notes the mean of the WPH statistics computed over the 9 separations done using different sub-maps (see Sect.~\ref{subsec:mock_data} for mock data construction). The colored bands represent $\pm 1\sigma$ error-bars, computed as the standard deviation of these statistics.}
    \label{fig:WPH_validation}
\end{figure}

\begin{figure}[h!]
    \centering
    \includegraphics[width=\hsize]{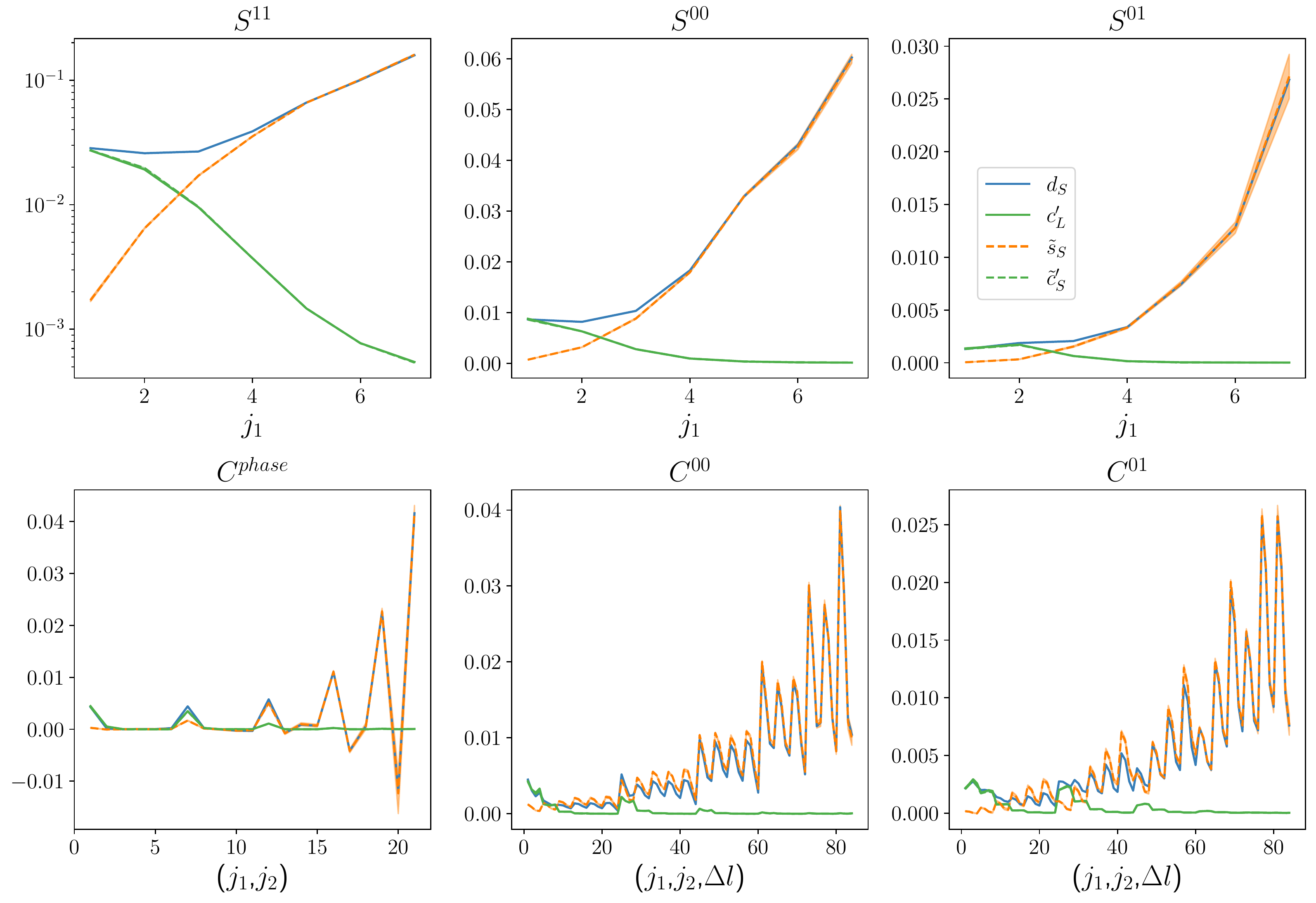}
    \caption{Same as Fig.~\ref{fig:WPH_validation} but for the \textit{Herschel} SPIRE maps at 250\micron. The WPH statistics of $d_S$, $\tilde{s}_S$, $c^{\prime}_{L}$, and $\tilde{c}_S^{\prime}$ are compared. The colored bands represent $\pm 1\sigma$ error-bars.}
    \label{fig:WPH_Herschel_separation}
\end{figure}

Here, we explain how to read the WPH statistics plots and we present the WPH statistics of the component separation results both on mock data (Fig.~\ref{fig:WPH_validation}) and on \textit{Herschel} observations (Fig.~\ref{fig:WPH_Herschel_separation}). These figures present the WPH statistics computed with $\Delta_n=0$ (no translation).
The values of $S^{11}$, $S^{00}$, and $S^{01}$, which characterize a dyadic scale and an angle, are averaged over the angles and plotted as a function of the scale, $j_1$, that they characterize. The $C^{\text{phase}}$ values, characterized by two scales and an angle, are averaged over the angles and plotted in lexicographic order as a function of $j_1$ and $j_2$. The $C^{00}$ and $C^{01}$ values, which are characterized by two scales and two angles, are averaged over the two angles at constant difference and plotted in lexicographic order as a function of $j_1 \in [0,J-1]$, $j_2 \in [j_1 +1,J-1]$ and $\Delta l \in [0,L-1]$ as the angle difference.

\section{(R)WST statistics}
\label{app:RWST}

The reduced wavelet scattering transform (RWST) statistics are interpretable non-Gaussian statistics defined in \citet{allys_rwst_2019}. They are a reduced form of the wavelet scattering transform (WST) statistics, first introduced in the field of data science \citep{mallat_group_2012}. This brief overview is directly inspired by \citet{allys_rwst_2019}.

\subsection{WST statistics}

The WST statistics are statistical descriptors that give a non-Gaussian description of a field. This is done by quantifying the level of coupling between scales. The WST are built by successive convolutions of the field with band-pass wavelets followed by the application of the modulus operator. Each of these summary statistics is then labeled by the scales they characterize. 

We consider a two-dimensional field named $\rho(\vec{x})$. As for the WPH statistics, the set of computed coefficients depends on two integers $J$ and $\Theta$, characterizing, respectively, the number of dyadic scales and angles considered. The integer scales $j$ are labelled from $0$ to $J-1$ and correspond to $2^j$ pixel scales, whereas the angles $\vartheta$ are labelled by integers $\theta \in [1,\Theta]$, such that $\vartheta = (\theta-1) \pi / \Theta$. The set of wavelets is then $\{ \psi_{j,\theta}(\vec{x}), j \in [0,J-1], \theta \in [1,\Theta] \}$. 

The WST coefficients are computed in three layers labelled by the integer $m$ going from 0 to 2. The only coefficient of the $m=0$ layer characterizes the average of the field:
\begin{equation}
    S_0 = \frac{1}{\mu_0} \int \rho(\vec{x}) \mathrm{d}^2 \Vec{x},
\end{equation}
where $\mu_0$ is the surface of integration. The coefficients $S_1(j_1,\theta_1)$ of the $m=1$ layer depend on a single oriented scale $(j_1,\theta_1)$ and are defined as:
\begin{equation}
    S_1(j_1,\theta_1) = \frac{1}{\mu_1} \int |\rho * \psi_{j_1,\theta_1}| (\Vec{x}) \mathrm{d}^2 \Vec{x},
\end{equation}
where $\mu_1$ is the impulse response
\begin{equation}
    \mu_1 = \int | \delta * \psi_{j_1,\theta_1}| (\Vec{x}) \mathrm{d}^2 \Vec{x},
\end{equation}
with $\delta$ the Dirac delta function. Finally, the coefficients $S_2(j_1,\theta_1,j_2,\theta_2)$ of the $m=2$ layer depend on two oriented scales $(j_1,\theta_1)$ and $(j_2,\theta_2)$, and are defined as:
\begin{equation}
    S_2(j_1,\theta_1,j_2,\theta_2) = \frac{1}{\mu_2} \int || \rho * \psi_{j_1,\theta_1}|*\psi_{j_2,\theta_2}|(\Vec{x}) \mathrm{d}^2 \Vec{x},
\end{equation}
where $\mu_2$ is defined in the same way as $\mu_1$. These coefficients are then normalized to separate the dependencies of the different layers. To this end, the coefficients of each layer are normalized by those of the previous layer. The normalized coefficients are denoted $\Bar{S}$ and defined as:
\begin{equation}
    \Bar{S}_1(j_1,\theta_1) = S_1(j_1,\theta_1) / S_0,
\end{equation}
and
\begin{equation}
\Bar{S}_2(j_1,\theta_1,j_2,\theta_2) = S_2(j_1,\theta_1,j_2,\theta_2) / S_1(j_1,\theta_1) , 
\end{equation}
whereas $\Bar{S}_0 = S_0$. These coefficients provide a rich non-Gaussian statistical description of non-Gaussian fields.

\subsection{RWST statistics}

The regularity of certain physical fields leads to possible simplifications of the WST statistics. The RWST allows for  the information contained in the WST to be concentrated into fewer coefficients, that are also more interpretable. This is done by fitting the angular dependencies of the logarithm of the WST coefficients with terms describing low-harmonics angular modulations. 

The only angular dependency of the $m=1$ layer coefficients is on $\theta_1$. The $\log_2 \left[ S_1 \right]$ coefficients are then written as:
\begin{equation}
\begin{split}
    \log_2 \left[ \Bar{S}_1(j_1,l_1) \right] = & \hat{S}_1^\text{iso}(j_1) + \\ 
    \hat{S}_1^\text{aniso}(j_1) & \times \cos \left( \frac{2 \pi}{\Theta} \left[ \theta_1 - \theta^{\text{ref},1}(j_1) \right] \right), \\
\end{split}
\end{equation}
where $\theta^{\text{ref},1}(j_1)$ is a reference angle describing the global anisotropy direction of the field. The $m=2$ layer is written using four terms and another reference angle:
\begin{equation}
\begin{split}
    \log_2 & \left[ \Bar{S}_2(j_1,\theta_1,j_2,\theta_2) \right] = \hat{S}_2^{\text{iso},1}(j_1,j_2)\\
    & + \hat{S}_2^{\text{iso},2}(j_1,j_2) \times \cos \left( \frac{2 \pi}{\Theta} \left[ \theta_1 - \theta_2 \right] \right) \\
    & + \hat{S}_2^{\text{aniso},1}(j_1,j_2) \times \cos \left( \frac{2 \pi}{\Theta} \left[ \theta_1 - \theta^{\text{ref},2}(j_1,j_2) \right] \right) \\
    & + \hat{S}_2^{\text{aniso},2}(j_1,j_2) \times \cos \left( \frac{2 \pi}{\Theta} \left[ \theta_2 - \theta^{\text{ref},2}(j_1,j_2) \right] \right). \\
\end{split}
\end{equation}
We refer to \citet{allys_rwst_2019} for an additional discussion on the interpretation of these coefficients.

\end{document}